# The 1916 PhD Thesis of Johannes Droste and the Discovery of Gravitational Repulsion

Charles H. McGruder III and B. Wieb VanDerMeer

Western Kentucky University

## Abstract

One of the biggest mysteries of astrophysics is the question of how highly energetic particles in relativistic jets and cosmic rays are accelerated. Recently, it has been suggested that gravitational repulsion is the mechanism responsible for these phenomena. The little known concept of gravitational repulsion was first introduced by Johannes Droste in his Ph.D. thesis submitted to H.A. Lorentz in 1916. It was written in Dutch. We provide a translation.

## Introduction

We observe highly energetic particles in the universe and we do not know how these particles are accelerated. We observe relativistic jets with speeds up to 0.8c (pulsar IGR J11014-6103) and apparent superluminal sources, which must have v > 0.707c. In addition we observe cosmic rays up to $10^{20}$ eV, whereby the most powerful man-made accelerator (Large Hadron Collider) can achieve proton-proton collisions of only $10^{12}$ eV. It has been suggested that the particles in relativistic jets (Gariel et al. 2016 and 2017) and cosmic ray particles (McGruder, 2017) are accelerated to high energies via gravitational repulsion. However, the concept of gravitational repulsion is not widely known even though as far back as 1916 Johannes Droste submitted a PhD thesis on general relativity to his advisor H.A. Lorentz in which he calculated the motion of a particle in what we call today the Schwarzschild field and found that highly relativistic particles experience gravitational repulsion. Thus, his thesis written in Dutch and never before translated contains the discovery of gravitational repulsion. Einstein was familiar with Droste's thesis, saying it is "good" (Schilling 2018).

In 1687 Newton published his universal law of gravitation, which states that the gravitational force is always attractive. This law is based on our terrestrial experience with slowly moving bodies ($v \ll c$). General Relativity (also referred to as Einstein's Theory of Gravitation) is valid not just for slowly moving bodies but also for those with relativistic velocities. After Einstein completed his formulation of General Relativity on 25 November 1915 (Janssen and Renn 2015), Johannes Droste on 8 December 1916 submitted his PhD thesis in Dutch containing the discovery that particles of sufficiently high velocity in a spherical symmetric gravitational field of a point mass according to Einstein's theory do not experience attraction rather repulsion. Fifteen days later on 23 December 1916 Hilbert submitted to the Society of Sciences in Göttingen a manuscript, which proved that Einstein's theory contains the concept of gravitational repulsion. Hilbert's work was published in 1917. Also in 1917 Droste published an abbreviated account of his thesis in English, which contained the concept of gravitational repulsion. In 1922 Bauer independently discovered this effect. Thus, Droste was the first to publish on gravitational repulsion in Einstein's theory of Gravitation and it is therefore historically incorrect to call the phenomenon "Hilbert Repulsion" as some authors do. (Loinger and Marsico 2009a, 2009b, 2010 and Célérier et al. 2017).

Below is a translation of Droste's entire thesis. Section 42 contains the discovery of gravitational repulsion.

# References

Bauer, H., Mathematische Einführung in die Gravitationstheorie Einsteins, Leipzig (1922)

Célérier M., Santos, N., Satheeshkumar, V., Hilbert repulsion in the Reissner-Nordström and Schwarzschild spacetimes, arXiv:1707.06994v1 [gr-qc], (2017)

Droste, j., Het zwaartekrachtsveld van een of meer lichamen volgens de theorie van Einstein, E. J. Brill, Leiden (1916)

Droste, J., Field of a single center in Einstein's theory of gravitation, and the motion of a particle in that field, Kon. Ak. Wetensch. Proc. **19**, 197 (1917)

Einstein, A., Die Feldgleichungen der Gravitation, Sitzungsberichte der Königlich Preussischen Akademie der Wissenschaften (Berlin), 844-847 (1915)

Gariel, J., Santos, N., Wang, A., Kerr geodesics following the axis of symmetry, Gen. Relativ. Gravit. **48**, id. 66, (2016)

Gariel, J., Santos, N. Wang, A., Observable acceleration of jets by a Kerr black hole, Gen. Relativ. Gravit. **49**, id 43 (2017)

Hilbert, D., Die Grundlagen der Physik (zweite Mitteilung), Nachrichten von der Gessellschaft der Wissenschaften zu Göttingen, Mathematisch-Physikalische Klasse 53 (1917)

Janssen, M., Renn, J., Arch and schaffold: How Einstein found his field equations, Phys. Today **68**, 30-36 (2015)

Loinger, A., Marsico, T., On Hilbert's Gravitational Repulsion (A Historical Note), arXiv:0904.1578 (2009a)

Loinger, A., Marsico, T., Protons do not exert any Hilbertian gravitational repulsion, arXiv:0912.1323 (2009b)

Loinger, A., Marsico, T., Gravitational collapses and Hilbertian repulsion, arXiv: 1008.3979 (2010)

McGruder III, C., Acceleration of particles to high energy via gravitational repulsion in the Schwarzschild field, Astropart. Phys. **86,** 18 (2017)

Schilling, G., Private communication (2018)

# THE GRAVITATIONAL FIELD OF ONE OR MORE BODIES ACCORDING TO EINSTEIN'S THEORY

## PH.D.THESIS

JOHANNES DROSTE, December 8, 1916

To the memory of my parents and for my uncle and aunt

At the completion of this thesis is it my desire to extend my thanks to all my professors and former professors of the college of Mathematics and Physics, whom's teaching I enjoyed, and to you, highly learned Bolland, for the part, that also Thou has played in my education.

In particular I thank you, very learned professor Kluyver and you, very learned professor Ehrenfest, for the interest, which you have shown in me so often.

Above everything, however, I think you, very learned professor Lorentz, esteemed mentor, not only for the many things you taught me during and after your lectures, but mainly for your cordial interest with which your followed my work and for your invaluable support, which you have given me during the preparation of this thesis. And the time I was allowed to be your assistant will always be for me a pleasant memory.

CONTENTS

INTRODUCTION

The principle of relativity, proposed by EINSTEIN in 1905 for systems moving at constant velocity, has developed itself within a few years, especially because of the work by EINSTEIN himself, to a general theory. The principle of equivalence between acceleration of a system and gravity opened up the the possibility to solve the problem of developing a theory of gravity. EINSTEIN accomplished this in 1913, but only a year ago EINSTEIN formulated equations for the gravitational field, which are covariant with all coordinate transformations. That his new theory can explain the precession of the perihelion of Mercury must be considered to be strong support for it.

In this thesis I formulate the calculation of the gravitational field of one single center at rest (chapter II) and the field of a number of centers, which are in motion with respect to each other (chapter III). The first calculation is exact and I have added to this calculation a discussion of the manner at which a mass point moves in that field; it will become clear how much more complicated already the calculation of this field is in EINSTEIN's theory compared to that in Newton's theory. The calculation of the field of a number of moving centers is an approximation, however it is more than sufficient for all applications in astronomy.

Preceding my calculation I have inserted a chapter in which briefly the main points of EINSTEIN's theory are explained (except for the part related to electromagnetism). This review will help the reader, who may not have EINSTEIN's publications handy, to reexamine the principles involved, and, it will serve as a source of equations, to which we can refer in both chapters that follow.

Finally I wish to point out, that I did not use anywhere the system of coordinates often used by EINSTEIN, in which $\sqrt{-g}=1$. Using this system does not simplify the calculations that much and it has, on the other hand, due to its unusual form, the disadvantage of focusing one's attention on properties of the coordinate system instead on characteristics of the equations.

# CHAPTER I

## REVIEW OF EINSTEIN'S GENERAL THEORY OF RELATIVITY

### § 1. **Introduction.**

1. Imagine a four-dimensional continuum; with this we mean the total of assigned values, which we can obtain by giving a value to each of four continuously variable quantities $x_1$, $x_2$, $x_3$, $x_4$. Such a set of values can be called a "point" in the continuum. It is clear, that the point ($x_1$, $x_2$, $x_3$, $x_4$) is not only known, if we know the values of $x_1$, $x_2$, $x_3$ and $x_4$, but also, if the values are known of four known functions $y_1$, $y_2$, $y_3$ and $y_4$ of $x_1$, $x_2$, $x_3$, $x_4$, provided that the equations

$$y_1 = f_1(x_1,x_2,x_3,x_4),\ y_2 = f_2(x_1,x_2,x_3,x_4), y_3 = f_3(x_1,x_2,x_3,x_4),\ y_4 = f_4(x_1,x_2,x_3,x_4)\ ..(1)$$

can be inverted in a unique way; and the quantities $y_1$, $y_2$, $y_3$ and $y_4$ will serve equally well for the description of the properties of the continuum as $x_1$, $x_2$, $x_3$ and $x_4$, if the functions

$$x_1 = \phi_1(y_1, y_2, y_3, y_4), \ x_2 = \phi_2(y_1, y_2, y_3, y_4), \ x_3 = \phi_3(y_1, y_2, y_3, y_4), \ x_4 = \phi_4(y_1, y_2, y_3, y_4),$$

which are obtained at this inversion, are continuous and also $f_1$, $f_2$, $f_3$ and $f_4$ are continuous.

2. The domain of observations exists in such a four-dimensional continuum. One of these four variables $x_1$, $x_2$, $x_3$, $x_4$, say $x_4$, can then be considered to represent the time. All "points" ($x_1$, $x_2$, $x_3$, $x_4$) for which $x_4$ has the same value, are the points in our normal three-dimensional space at the moment given by the value of $x_4$. Those points can be called *simultaneous* in the four-dimensional space. On the other hand, one can say that two points ($x_1$, $x_2$, $x_3$, $x_4$) and ($x_1$, $x_2$, $x_3$, $x_4$' ) represent the *same point* in normal space at different moments in time. When one now transforms from the system of variables $x_1$, $x_2$, $x_3$, $x_4$ to the system $y_1$, $y_2$, $y_3$, $y_4$, and, taking $y_4$ as the time, one makes a corresponding group of points in four-dimensional space as simultaneous, then this group will be different, because the points $x_4$ = constant will contain different $y_4$ values and the points $y_4$ = constant will also contain different $x_4$ values. The points ($x_1$, $x_2$, $x_3$, $x_4$) and ($x_1$, $x_2$, $x_3$, $x_4$' ), which in the first group represented the same three-dimensional point in space, will not be considered to be that in the second group, because at unchanged values of $x_1$, $x_2$ and $x_3$ but changed value of $x_4$, all the quantities $y_1$, $y_2$, $y_3$ and $y_4$, will have obtained different values. Only if the transformation is of the form

$$y_1 = f_1(x_1, x_2, x_3), \quad y_2 = f_2(x_1, x_2, x_3), \quad y_3 = f_3(x_1, x_2, x_3), \quad y_4 = f_4(x_4),$$

will the identity of the points of the (normal) space and the simultaneousness not be affected by the transformation. In that case one has a normal coordinate transformation and a separate time transformation.

The principle of EINSTEIN's general theory of relativity is then, that the different choices for variables are equivalent, so that "simultaneousness" and "identity of points in space" have only meaning with respect to a "coordinate system" ($x_1$, $x_2$, $x_3$, $x_4$) chosen beforehand. On the other hand, one can consider the points in four-dimensional space to be independent of the coordinate system, insofar as one can regard the set of values $x_1$, $x_2$, $x_3$, $x_4$ and the set $y_1$, $y_2$, $y_3$, $y_4$ as the same point as the last set is obtained from the first set according to equation 1. The principle of relativity will be formulated in point 5 below in more detail.

3. A quantitative relation between different points in the four-dimensional space is herewith not yet proposed. Because a numerical value related to two points ($x_1$, $x_2$, $x_3$, $x_4$) and ($x_1'$, $x_2'$, $x_3'$, $x_4'$), for example $(x_1-x_1')^2 + (x_2-x_2')^2 + (x_3-x_3')^2 + (x_4-x_4')^2$ will not be equal to the corresponding value $(y_1-y_1')^2 + (y_2-y_2')^2 + (y_3-y_3')^2 + (y_4-y_4')^2$, which belongs to the "same" points ($y_1$, $y_2$, $y_3$, $y_4$) and ($y_1'$, $y_2'$, $y_3'$, $y_4'$). To define a "distance" between two "points" one writes an expression, which depends on the coordinates of both points in the same way and, which due to a convenient convention about the way it transforms, keeps its form. The usual distance

$$d\,x_1^2 + d\,x_2^2 + d\,x_3^2 + d\,x_4^2$$

between two infinitely close points ($x_1$, $x_2$, $x_3$, $x_4$) and ($x_1+dx_1$, $x_2+dx_2$, $x_3+dx_3$, $x_4+dx_4$) obtains with the transformation in equation (1) the form

$$\sum_{ij} a_{ij} dy_i dy_j ,$$

of which the original form a special case is; $a_{ij}$ are functions $y_1$, $y_2$, $y_3$ and $y_4$. As a result, the "distance" between two infinitely little different points is

$$ds^2 = \sum_{ij} g_{ij} dx_i dx_j \quad \text{...............................................(2)}$$

Performing the transformation in equation (1), results in

$$dx_i = \sum_l p_{il} dy_l \quad \text{....................................................(3)}$$

if

$$p_{il} = \frac{\partial x_i}{\partial y_l}.$$

Therefore one obtains

$$ds^2 = \sum_{ijkl} g_{ij} p_{il} p_{jk} dy_l dy_k = \sum_{lk} \left( \sum_{ij} g_{ij} p_{jl} p_{jk} \right) dy_l dy_k$$

and, if one now poses

$$g'_{lk} = \sum_{ij} g_{ij} p_{li} p_{jk} \quad \text{...............................................(4)}$$

then we have

$$ds^2 = \sum_{lk} g'_{lk} dy_l dy_k,$$

which has the same form as equation (2). Thus, by choosing (2) as the definition of the distance and by the prescription, that in another coordinate system $y_1$, $y_2$, $y_3$, $y_4$ the quantities $g_{ij}$ must be replaced by the quantities $g'_{lk}$, calculated from $g_{ij}$ according to equation (4), one finds, that $ds^2$ in all systems is represented by an expression of the same form.

In equation (2) both $g_{ij}dx_i dx_j$ and $g_{ji}dx_j dx_i$ will appear if $i$ and $j$ are different indices. These terms together yield $(g_{ij} + g_{ji})dx_i dx_j$, so that neither $g_{ij}$ nor $g_{ji}$ is important, but the sum $g_{ij} + g_{ji}$ is relevant. By redefining $g_{ij}$ and $g_{ji}$ as $g_{ij} = \frac{1}{2}(g_{ij} + g_{ji})$, so that $g_{ij}$ and $g_{ji}$ become equal to each other, and also $g'_{ij}$ and $g'_{ji}$ become equal to each other as follows from equation (4).

4. The transformation in equation (1) can be determined in such a way, that the functions $g'_{lk}$ is a specific point, for example in ($x_1^0$, $x_2^0$, $x_3^0$, $x_4^0$) become equal to zero, if $l \neq k$ is and $\pm 1$ for equal values of $l$ and $k$. Then we have in that point

$$ds^2 = \pm dy_1^2 \pm dy_2^2 \pm dy_3^2 \pm dy_4^2 \quad\quad\quad\quad\quad\quad\quad\quad\quad\quad (5)$$

and because the functions $g'_{ik}$ are continuous functions of the coordinates (we will even require in § 3 and 4, that they can be differentiated twice) this expression will be valid by approximation in a relatively large area near ($x_1^0$, $x_2^0$, $x_3^0$, $x_4^0$). In all linear orthogonal transformations, which can be applied to the system of y's, $ds^2$ will keep that form there. EINSTEIN connects the theory here to the special theory of relativity, by presuming that the expression for $ds^2$, given by equation (5), represents the following invariant differential form in the special theory of relativity:

$$dt^2 - dx^2 - dy^2 - dz^2 \;.$$

If one imagines then in each point ($x_1$, $x_2$, $x_3$, $x_4$) the existence of the quadratic expression

$$\sum_{ij} g_{ij} \xi_i \xi_j = \varepsilon^2$$

in which $\xi_i$ represent running coordinates ($g_{ij}$ remains thereby constant) and $\varepsilon$ is a constant, then EINSTEIN requires, that this expression has a real and three imaginary axes. We will call this the "indicatrix" in the point ($x_1$, $x_2$, $x_3$, $x_4$). If one assigns the length $\varepsilon$ to the line segment connecting the center ($x_1$, $x_2$, $x_3$, $x_4$) of the indicatrix to one of its points ($x_1+\xi_1$, $x_2+\xi_2$, $x_3+\xi_3$, $x_4+\xi_4$), and if one determines that a point ($x_1+dx_1$, $x_2+dx_2$, $x_3+dx_3$, $x_4+dx_4$), situated on that line segment, a distance to ($x_1$, $x_2$, $x_3$, $x_4$) has in proportional to $\varepsilon$ as $dx_1:\xi_1 = dx_2:\xi_2 = dx_3:\xi_3 = dx_4:\xi_4$, then this distance becomes exactly equal to $ds$. One calls $ds$ the *natural distance* and can say, therefore, that the line segment divided by $\varepsilon$ the *natural measure* of the indicatrix is. It is clear, that for those points, for which the line segment does not intersect with the indicatrix, the distance becomes imaginary. For specifics see "On EINSTEIN's theory of gravity", by H.A. LORENTZ (in Dutch), I, Zittingsversl. Kon. Akad. V. Wetensch., Amesterdam, part 24, page 1389.

    5. The general theory of relativity requires now, that the description of phenomena can happen using such quantities, that the equations, which describe them, keep their form under an arbitrary coordinate transformation. This will then considered to be sufficient to ensure, that each phenomenon, which happens in one coordinate system in a certain way, corresponds with another phenomenon, that in an other coordinate system will happen exactly in the same way as the first phenomenon in the first system.

    To ensure this we will consider all quantities to be geometric quantities; such quantities transform via a coordinate transformation in a certain way. Before we can write down the equations proposed by EINSTEIN for the general theory of relativity, we must first discuss the most important geometric quantities.

## § 2. **Geometric quantities**.

6. Performing the coordinate transformation in equation (1), we propose

$$p_{il} = \frac{\partial x_i}{\partial y_l}, \quad \pi_{li} = \frac{\partial y_i}{\partial x_l},$$

Equation (3) holds then and the inverse of that equation:

$$dx_i = \sum_l p_{il} dy_l, \quad dy_i = \sum_l \pi_{li} dx_l \quad (i = 1,2,3,4),$$

The quantities $g_{ij}$ transform according to equation (4). If $g^{ij}$ represent the algebraic complements of the quantities $g_{ij}$, that are quantities for which holds:

$$\sum_j g_{ij} g^{jl} = 0 \, (i \neq l), \sum_j g_{lj} g^{jl} = 1.$$

One then finds easily, that under the transformation of equation (1) these quantities transform into $g^{ij}{}'$, according to:

$$g^{lk}{}' = \sum_{ij} \pi_{il} \pi_{jk} g^{ij}.$$

All quantities, which at a coordinate transformation change like $dx_i, g_{ij}, g^{ij}$ into linear functions of the original values with coefficients, which consist of productions of $p$'s, $\pi$'s or both, are called tensors. RICCI and LEVI-CIVITÀ proposed the theory for tensors.

*Definitions.* 1. A *covariant tensor* of rank $\lambda$ is a system of functions $T_{i_1 i_2 \ldots i_\lambda}$ of $x_1, x_2, x_3, x_4$, which transform according to the formulas

$$T_{j_1 j_2 \ldots j_\lambda}' = \sum_{i_1 i_2 \ldots i_\lambda} p_{i_1 j_1} p_{i_2 j_2} \ldots p_{i_\lambda j_\lambda} T_{i_1 i_2 \ldots i_\lambda},$$

2. A *contravariant tensor* of rank $\lambda$ is a system of functions $T^{i_1 i_2 \ldots i_\lambda}$ of $x_1, x_2, x_3, x_4$, which transform according to the formulas

$$T^{j_1 j_2 \ldots j_\lambda}{}' = \sum_{i_1 i_2 \ldots i_\lambda} \pi_{i_1 j_1} \pi_{i_2 j_2} \ldots \pi_{i_\lambda j_\lambda} T^{i_1 i_2 \ldots i_\lambda}{}',$$

3. A *mixed tensor* of rank $\lambda$ and contravariant tensor of rank $\mu$, is a system of functions $T^{j_1 j_2 \ldots j_\mu}_{i_1 i_2 \ldots i_\lambda}$ of $x_1, x_2, x_3, x_4$, which transform according to the formulas

$$T^{k_1 k_2 \ldots k_\mu}_{l_1 l_2 \ldots l_\lambda}{}' = \sum_{i_1 i_2 \ldots i_\lambda}\sum_{j_1 j_2 \ldots j_\mu} p_{i_1 l_1} p_{i_2 l_2} \ldots p_{i_\lambda l_\lambda} \pi_{j_1 k_1} \pi_{j_2 k_2} \ldots \pi_{j_\mu k_\mu} T^{j_1 j_2 \ldots j_\mu}_{i_1 i_2 \ldots i_\lambda},$$

(Translators note: in the original the second p's subscript was *i₂i₂*, which was probably a typo. It is changes to *i₂l₂*).

From these definitions it is clear that $dx_i$ a contravariante tensor of the first rank is. Tensors of the first rank are also called vectors. $g_{ij}$ and $g^{ij}$ are tensors of the second rank; $g_{ij}$ is covariant and $g^{ij}$ is contravariant. $g_{ij}$ is sometimes called the *fundamental covariant tensor* and $g^{ij}$ the *fundamental contravariant tensor*.

A single quantity *T,* which in each coordinate system has the same value, is a tensor (co-or contra- as one likes) of rank zero. It is called a scalar quantity or simply a scalar.

One can prove easily, that if, for example, $A_{ijl}$ and $B^{ij}$ are a covariant and a contravariant tensor, then

$$\sum_j A_{ijl} B^{kj}, \sum_{ij} A_{ijl} B^{kj}, A_{ijl} B^{kn}$$

represents tensors; the first is covariant of the second rank and contravariant of the first rank; the second is a covariant vector; the third is covariant of the third rank and contravariant of the second rank. So is, for example,

$$\delta_i^j = \sum_l g_{il} g^{ij} = \begin{cases} 1, \text{for } j = i \\ 0, \text{for } j \neq i \end{cases} \quad \text{............(6)}$$

a mixed tensor, co- and contravariant of the first rank. For an extensive discussion of such constructs one should consult A. EINSTEIN, "Die formale Grundlage des allgemeinen Relativitätstheorie" Sitzungsver. de Kön. Preuss. Akad. d. Wiss. 1914, page 1030.

7. The determinant of the quantities $g_{ij}$ can be called $g$. It is always negative. The subdeterminants divided by $g$ itself, if given a suitable sign, are the quantities $g^{ij}$ introduced in section 1 above. One can see easily, using the product rule of determinants, that via a transformation we obtain

$$g' = p^2 g,$$

if $p$ represents the functional-determinant $\begin{pmatrix} x_1 & x_2 & x_3 & x_4 \\ y_1 & y_2 & y_3 & y_4 \end{pmatrix}$.

From the known rule for the transformation of a four-dimensional integral follows easily, that the following equality holds:

$$\iiint\int \sqrt{-g}\, dx_1\, dx_2\, dx_3\, dx_4 = \iiint\int \sqrt{-g'}\, dy_1\, dy_2\, dy_3\, dy_4\ ;$$

the integration is thereby extended over a arbitrary area.

8. CHRISTOFFEL, to whom we owe the theory of quadratic differential forms, introduces the following notation:

$$\begin{bmatrix} i\ j \\ l \end{bmatrix} = \tfrac{1}{2}\left( \frac{\partial g_{il}}{\partial x_j} + \frac{\partial g_{jl}}{\partial x_i} - \frac{\partial g_{ij}}{\partial x_l} \right),\ \begin{Bmatrix} i\ j \\ l \end{Bmatrix} = \sum_k \begin{bmatrix} i\ j \\ l \end{bmatrix} g^{kl}.$$

These are called the symbols of CHRISTOFFEL of the first and second type. If $T_{i_1 i_2 \cdots i_\lambda}$ is a covariant tensor is, one can prove that

$$T_{i_1 i_2 \ldots i_\lambda j} = \frac{\partial T_{i_1 i_2 \ldots i_\lambda}}{\partial x_j} - \sum_k \left( \begin{Bmatrix} i_1\ j \\ k \end{Bmatrix} T_{k i_2 \ldots i_\lambda} + \begin{Bmatrix} i_2\ j \\ k \end{Bmatrix} T_{i_1 k \ldots i_\lambda} + \ldots + \begin{Bmatrix} i_\lambda\ j \\ k \end{Bmatrix} T_{i_1 i_2 \ldots k} \right)$$

is a tensor of rank $\lambda + 1$. It is said, that it is obtained from $T_{i_1 i_2 \cdots i_\lambda}$ by *covariant differentiation*.

If $T^{i_1 i_2 \cdots i_\lambda}$ is a contravariant tensor, then

$$T^{i_1 i_2 \ldots i_\lambda j} = \sum_l g^{jl} \frac{\partial T^{i_1 i_2 \ldots i_\lambda}}{\partial x_l} + \sum_{kl} g^{jl}\left( \begin{Bmatrix} l\ k \\ i_1 \end{Bmatrix} T^{k i_2 \ldots i_\lambda} + \begin{Bmatrix} l\ k \\ i_2 \end{Bmatrix} T^{i_1 k \ldots i_\lambda} + \ldots + \begin{Bmatrix} l\ k \\ i_\lambda \end{Bmatrix} T^{i_1 i_2 \ldots k} \right)$$

is a contravariant tensor of rank $\lambda + 1$, obtained by *contravariant differentiation* from $T^{i_1 i_2 \cdots i_\lambda}$.

9. When one differentiates $g_{ij}$ covariantly and $g^{ij}$ contravariantly, then one obtains zero identically. There is, however, a tensor of the fourth rank, which is formed from the fundamental tensor. It is of great importance for the question, whether two quadratic differential forms can transform into each other. One poses

$$(i\,j,k\,l) = \tfrac{1}{2}\left(\frac{\partial^2 g_{il}}{\partial x_j \partial x_k} + \frac{\partial^2 g_{jk}}{\partial x_i \partial x_l} - \frac{\partial^2 g_{ik}}{\partial x_j \partial x_l} - \frac{\partial^2 g_{lj}}{\partial x_i \partial x_k}\right) + \sum_{mn} g^{mn}\left(\begin{bmatrix} i\,l \\ m \end{bmatrix}\begin{bmatrix} j\,k \\ n \end{bmatrix} - \begin{bmatrix} i\,k \\ m \end{bmatrix}\begin{bmatrix} j\,l \\ n \end{bmatrix}\right)$$

and calls $(i\,j,k\,l)$ a symbol of RIEMANN. The system of quantities

$$R_{ijkl} = (i\,j,k\,l)$$

is a tensor of the fourth rank. The fact that this tensor is identically zero is, as one has proven, necessary and sufficient to be able to transform $ds^2$ to the form $\sum_i \pm dy_i^2$. From the tensor $(i\,j,k\,l)$ one can derive the tensor $\{i\,j,k\,l\}$ using the formulas

$$\{i\,j,k\,l\} = \sum_n g^{jn}(i\,n,k\,l),$$

which is covariant of the third rank and contravariant of the first rank.
For $(i\,j,k\,l)$ one has

$$\{i\,j,k\,l\} = \frac{\partial \left\{\begin{matrix} i\,k \\ j \end{matrix}\right\}}{\partial x_l} - \frac{\partial \left\{\begin{matrix} i\,l \\ j \end{matrix}\right\}}{\partial x_k} + \sum_n \left(\left\{\begin{matrix} i\,k \\ n \end{matrix}\right\}\left\{\begin{matrix} n\,l \\ j \end{matrix}\right\} - \left\{\begin{matrix} i\,l \\ n \end{matrix}\right\}\left\{\begin{matrix} n\,k \\ j \end{matrix}\right\}\right),$$

The tensor

$$G_{il} = \sum_k \{i\,k,k\,l\}$$

is covariant of the second rank; the scalar

$$G = \sum_{il} G_{il} g^{il} = \sum_{ilk} g^{il} \{i\ k, k\ l\}$$

is called the *curvature* of the extensiveness measured by the line-element $ds^2 = \sum_{ij} g_{ij} dx_i dx_j$ .

### § 3. The equations of the general theory of relativity.

10. When in the four-dimensional space a line is given, then one can consider the points of this line to represent the same point at different moments in time, or not, depending on the choice of coordinates $x_1, x_2, x_3, x_4$ . In the first case the line represents a stationary point, in the second case a moving point; in both cases the point is called a *world line*.

World lines can be points of moving bodies or light rays. World lines refer to all phenomena for which one can assume that what happens now here is identical to what happens soon over there. The world lines of light rays and those of mass points are immediately given by the observations in space. EINSTEIN requires:

1. The lines for light are minimal lines, that is, lines for which holds

$$ds^2 = 0.$$

2. World lines of mass point are the geodetic lines, that is, lines, for which the length

$$\int ds$$

between two fixed points changes infinitely little of higher order (than the first), if one varies the in-between points over distances that are of the first order, while keeping the end points unchanged,

11. *Light rays.* If one chooses a coordinate system $x_1, x_2, x_3, x_4$, and in the point $(x_1, x_2, x_3)$ at the moment $x_4$ a certain direction is given by the ratios $dx_1 : dx_2 : dx_3$, then the equation

$$\sum_{ij} g_{ij} dx_i dx_j = 0$$

will give in general two values for $dx_4 : dx_1$, which will depend on the values of $x_1, x_2, x_3$ and $x_4$; if they differ in sign, the light can propagate in both directions along the element $(dx_1, dx_2, dx_3)$. Both values correspond to propagation in of the two directions given by the two possible ratios $dx_1 : dx_2 : dx_3$. We find, therefore, that light cannot propagate in all directions, not even in two opposite directions, with the same speed, not withstanding that the speed will depend on position and time. Yet, these velocities depend on the coordinate system; if one introduces the system described in point 4 above, so that $ds^2$ near ($x_1^0$, $x_2^0$, $x_3^0$, $x_4^0$) will take the form

$$ds^2 = dy_4^2 - dy_1^2 - dy_2^2 - dy_3^2,$$

then all those special properties are lifted *locally* and *temporarily*. Lifting them everywhere and at always is not possible, unless the from $\sum_{ij} g_{ij} dx_i dx_j$ can be transformed into $dy_4^2 - dy_1^2 - dy_2^2 - dy_3^2$. And, this is only the case, if $(i\,j, k\,l) = 0$ for all combinations of the indices. If the symbols of RIEMANN are not equal to zero, then the anisotropy in the propagation of light is, therefore, an essential property of space.

The equations, which EINSTEIN poses for the electromagnetic field and which not only determine the speed, at which a line element will be crossed, but also the whole propagation of light, are in agreement with the condition $ds^2 = 0$.

12. *Mass point.* Along the world line for a mass point are three of the coordinates functions of the fourth; is is also possible to consider each of the four coordinates as a function of a parameter $\lambda$. The condition

$$\delta \int_{\lambda_1}^{\lambda_2} \sqrt{\sum_{ij} g_{ij} \frac{dx_i}{d\lambda} \frac{dx_j}{d\lambda}} \, d\lambda = 0$$

yields the equations

$$\frac{d}{d\lambda}\left( \frac{1}{L} \sum_j g_{ij} \frac{dx_j}{d\lambda} \right) - \frac{1}{2L} \sum_{jl} \frac{\partial g_{jl}}{\partial x_i} \frac{dx_l}{d\lambda} \frac{dx_j}{d\lambda} = 0 \qquad \text{............(7)}$$

in which $L = \sqrt{\sum_{ij} g_{ij} \frac{dx_i}{d\lambda} \frac{dx_j}{d\lambda}}$ is.

These equations are simplest, if one eliminates $\lambda$ from (7) and introduces a new independent variable, the value $s$ of $\int ds$, taken from a previously determind point on the world line up to the point, for which the value $s$ will refer to. Then $L$ will cancel out of the formulas and one finds easily, by evaluating the differentiation in the first term,

$$\frac{d^2 x_i}{ds^2} + \sum_{jl} \left\{ \begin{array}{c} j\,l \\ i \end{array} \right\} \frac{dx_l}{ds} \frac{dx_j}{ds} = 0 \qquad \text{,........................(8)}$$

If $g_{ij} = 0$, for $i \neq j$, and if $g_{11} = g_{22} = g_{33} = -1$ and $g_{44} = 1$, then the world line of each mass point is straight and this means, that it moves uniformly along a straight line. In other cases the movement is different and this relates according to (7) to the differential quotients of the $g$'s to the coordinates; for these reasons the $g$'s are also called *gravitational potentials*.

13. In the special theory of relativity the three equations of motion and the energy equation in the case of a continuously distributed mass have the form

$$\left.\begin{aligned}\frac{\partial p_{xx}}{\partial x}+\frac{\partial p_{xy}}{\partial y}+\frac{\partial p_{xz}}{\partial z}+\frac{\partial i_x}{\partial t}&=f_x,\\ \frac{\partial p_{yx}}{\partial x}+\frac{\partial p_{yy}}{\partial y}+\frac{\partial p_{yz}}{\partial z}+\frac{\partial i_y}{\partial t}&=f_y,\\ \frac{\partial p_{zx}}{\partial x}+\frac{\partial p_{zy}}{\partial y}+\frac{\partial p_{zz}}{\partial z}+\frac{\partial i_z}{\partial t}&=f_z,\\ \frac{\partial s_x}{\partial x}+\frac{\partial s_y}{\partial y}+\frac{\partial s_z}{\partial z}+\frac{\partial \varepsilon}{\partial t}&=w.\end{aligned}\right\} \quad \ldots(9)$$

$p_{xx}$, $p_{xy}$, etc. represent the internal stresses ($p_{xy}=p_{yx}$ etc.), $i_x$, $i_y$, $i_z$ are the components of the momentum per unit of volume, $s_x, s_y, s_z$ are the components of the energy current, $\varepsilon$ is the energy per unit volume, $f_x, f_y, f_z$ are the components of the external force per unit of volume and $w$ is the added energy per unit of volume and unit of time. The equations speak for themselves. In the special theory of relativity it the vector $(i_x, i_y, i_z)$ equal to the vector $(s_x, s_y, s_z)$ apart from a constant factor, which dpends on the speed of light and which can be set equal to 1 by changing the choice of units.

These equations correspond to the following equations in the general theory of relativity

$$\sum_j \frac{\partial}{\partial x_j}\left(\sqrt{-g}T_i^{\ j}\right) = \tfrac{1}{2}\sum_{jlk}\sqrt{-g}g^{kl}\frac{\partial g_{ij}}{\partial x_i}T_k^{\ j}+K_i\left(i=1,2,3,4\right) \quad ..(10)$$

in which $T_i^{\,j}$ is a mixed tensor of the second rank. (Translator's note: the original manuscript had a typo in the line following this note, but a correction was inserted at the introduction. This correction has been executed in this very line). If one poses

$$T_{ij} = \sum_l g_{jl} T_i^{\,l} \text{ and } T^{ij} = \sum_l g^{il} T_l^{\,j},$$

one can also write

$$\sum_{jl} \frac{\partial}{\partial x_j}\left(\sqrt{-g}\, g^{lj} T_{il}\right) = \tfrac{1}{2} \sum_{jlkn} \sqrt{-g}\, g^{kl} g^{nj} \frac{\partial g_{ij}}{\partial x_i} T_{kn} + K_i \, (i=1,2,3,4) \quad ..(11)$$

and

$$\sum_{lj} \frac{\partial}{\partial x_j}\left(\sqrt{-g}\, g_{li} T^{\,jl}\right) = \tfrac{1}{2} \sum_{jl} \sqrt{-g}\, \frac{\partial g_{lj}}{\partial x_i} T^{\,jl} + K_i \, (i=1,2,3,4) \quad ...........(12)$$

Each of these sets of four equations keeps its form under a coordinate transformation. Because, if one transfers the first term of the second part to the first part and divides the equation by $\sqrt{-g}$, then the first part becomes a covariant vector, actually that covariant vector, that one gets by differentiating the tensor $T_{il}$ covariantly, so that one generates $T_{ilj}$, and, after that to multiply with $g^{lj}$ and to sum over the indices $l$ and $j$, Therefore, $K_i/\sqrt{-g}$ is also a covariant vector.

If

$$ds^2 = -dx_1^2 - dx_2^2 - dx_3^2 + dx_4^2,$$

then (10) becomes

$$\sum_j \frac{\partial}{\partial x_j}\left(\sqrt{-g}\, T_i^{\,j}\right) = K_i, \quad (i=1,2,3,4) \quad ,................(13)$$

and this will agree with (9). EINSTEIN supposes, that in this case we have

$$\sqrt{-g}T_1^1 = -p_{xx}, \sqrt{-g}T_1^2 = -p_{xy}, \text{ etc.}, \sqrt{-g}T_1^4 = -i_x, \text{ etc.}, \sqrt{-g}T_4^1 = s_x, \text{ etc.},$$

$$\sqrt{-g}T_4^4 = \varepsilon \quad\text{...........................................................}(14),$$

so that the first of the three equations in (13) become the first three of (9) by multiplying with -1, and the fourth equation in (13) transforms immediately into the fourth of (9).

Finally, we would like to point out, that in the case of a cluster of mass points, which do not influence each other, and which move in such a way, that the velocity components are continuous functions of $x_1, x_2, x_3$ and $x_4$, we have

$$T^{ij} = \rho \frac{dx_i}{ds}\frac{dx_j}{ds},$$

where $\rho$ is a scalar and represents the "density at rest" ("Ruhdichte"). From here and from (12) one can derive (8). Inversely, it is possible to derive (12) from the variation principle in point 12. For this last statement see H.A. LORENTZ "Het beginsel van HAMILTON in EINSTEIN's theorie der zwaartekracht" Zittingsversl. Kon. Akad. v. Wetensch., Deel XXIII, page 1073.

14. What remains is to answer the question, how the gravitation field will behave at a given mass distribution. EINSTEIN gives formulas, from where the $g$'s can be calculated, when the $T$'s are given. We will call those formulas the equations of gravity. They take the place of the formula

$$\Delta \varphi = \rho$$

in NEWTON's theory. The tensor $G_{ij}$ defined in point 9 plays a role here. EINSTEIN supposes

$$G_{ij} = -\kappa\left(T_{ij} - \tfrac{1}{2}g_{ij}T\right), \quad (i,j=1,2,3,4) \quad\text{.........}(15)$$

In this equation

$$T = \sum_{ij} g^{ij} T_{ij}$$

is a scalar and $\kappa$ is a constant, the gravitational constant. One can also give this equation the form

$$G_{ij} - \tfrac{1}{2} g_{ij} G = -\kappa T_{ij}, \quad (i,j = 1,2,3,4) \quad \ldots\ldots(16)$$

Both sides represent a covariant tensor; so the equations are covaraint under all coordinate transformations.

     15. EINSTEIN succeeded also to describe the electromagnetic phenomena using completely covariant equations. We will not discuss that, however. We will also not describe the nice form, that LORENTZ has given to EINSTEIN's theory, at which that part of the theory we have discussed here is represented without coordinates (Over EINSTEIN's theorie der zwaartekracht" Zittingsversl. Kon. Akad. v. Wetensch., Deel XXIV, I page 1389, II page 1759). We will also pass by the interesting work of HILBERT (Göttinger Nachrichten, Math-phys. Kl. 1915, page 395) in silence. I wanted to give in the above only so much as is strictly necessary for the understanding of my following calculations. However, I must mention the variation theorem (see Hilbert at the reference above), from which the equations (16) can be derived. This is

$$\delta \int\int\int\int G\sqrt{-g}\, dx_1\, dx_2\, dx_3\, dx_4 = \int\int\int\int \sum_{ij} T_{ij} \delta g^{ij} dx_1 dx_2 dx_3 dx_4 \quad \ldots\ldots(17).$$

Each of the integrals in this equation must be extended over a four-dimensional area, and the variations of the $g$'s must be taken in such a way, that they are equal to zero at the (three-dimensional) border of that area; the same is true for the variations of the first derivatives of the $g$'s with respect to the coordinates.

# CHAPTER II.

## THE FIELD OF A SINGLE SPHERICAL CENTER

### § 1. **Calculation of the field**.

16. We are going to search for an expression of $ds^2$, that is suitable to represent the gravitational field of a single spherical center. We imagine this center to be at rest; the $g$'s will therefore be independent of time. As for the other three coordinates, it should be possible to choose these in such a way, that the $g$'s will only depend on one of them, which we will call $r$, while the other two, the polar coordinates $\vartheta$ and $\varphi$, will appear in $ds^2$ only in the form $d\vartheta^2 + \sin^2 \vartheta d\varphi^2$, so that we have achieved that $ds^2$ keeps the form at any other choice of the direction $\vartheta = 0$. We pose, therefore,

$$ds^2 = w^2 dt^2 - u^2 dr^2 - v^2 \left( d\vartheta^2 + \sin^2 d\varphi^2 \right) \quad \text{................(18)}$$

By also choosing the coefficients of $drdt$, $d\vartheta dt$ and $d\varphi dt$ equal to zero, we find that $ds^2$ by the substitution $t' = -t$ will remain unchanged and the motions in the field will be reversible.

17. In order to calculate the field, we use the variation theorem, with which chapter I concludes and which outside the center itself assumes the form

$$\delta \int\int\int\int G\sqrt{-g}\, dx_1 dx_2 dx_3 dx_4 = 0 \quad \text{................(19)}$$

It is therefore important to calculate $G$ first.

We will show that, as soon $g_{ij}$ (and therefore also $g^{ij}$) are equal to zero in an area for *different* values for both indices $i$ and $j$, $K$ breaks up in six pieces, which each relate to one of the six combinations of two indices.

Let $a, b$ and $c$ represent three different indices. One has then

$$\left[\begin{array}{c} a\ b \\ c \end{array}\right] = 0, \left[\begin{array}{c} a\ a \\ c \end{array}\right] = -\tfrac{1}{2}\frac{\partial g_{aa}}{\partial x_c}, \left[\begin{array}{c} a\ b \\ a \end{array}\right] = \tfrac{1}{2}\frac{\partial g_{aa}}{\partial x_b}, \left[\begin{array}{c} a\ a \\ a \end{array}\right] = \tfrac{1}{2}\frac{\partial g_{aa}}{\partial x_a}$$

and from there it follows

$$\left\{\begin{array}{c} a\ b \\ c \end{array}\right\} = 0, \left\{\begin{array}{c} a\ a \\ c \end{array}\right\} = -\tfrac{1}{2}g^{cc}\frac{\partial g_{aa}}{\partial x_c}, \left\{\begin{array}{c} a\ b \\ a \end{array}\right\} = \tfrac{1}{2}g^{aa}\frac{\partial g_{aa}}{\partial x_b},$$

$$\left\{\begin{array}{c} a\ a \\ a \end{array}\right\} = \tfrac{1}{2}g^{aa}\frac{\partial g^{aa}}{\partial x_a} \quad\text{...........................................(20)}$$

In the expression

$$2G = 2\sum_{ij} g^{ii}\left(\frac{\partial}{\partial x_i}\left\{\begin{array}{c} i\ j \\ j \end{array}\right\} - \frac{\partial}{\partial x_j}\left\{\begin{array}{c} i\ i \\ j \end{array}\right\}\right) + 2\sum_{ijk} g^{ii}\left(\left\{\begin{array}{c} i\ j \\ k \end{array}\right\}\left\{\begin{array}{c} k\ i \\ j \end{array}\right\} - \left\{\begin{array}{c} i\ i \\ k \end{array}\right\}\left\{\begin{array}{c} k\ j \\ j \end{array}\right\}\right)$$

we now consider the first term

$$2G_1 = 2\sum_{ij} g^{ii}\left(\frac{\partial}{\partial x_i}\left\{\begin{array}{c} i\ j \\ j \end{array}\right\} - \frac{\partial}{\partial x_j}\left\{\begin{array}{c} i\ i \\ j \end{array}\right\}\right).$$

Only in $G_i$ appear second derivatives of the $g$'s with respect to the coordinates. If we now realize, that $i = j$ in $G_i$ yields *zero*, then we see, that $2G_i$ can be split up in 6 parts, each related to one of the 6 combinations two-to-two of the four indices 1,2,3,4. The part, that refers to $\alpha\beta$, is obtained by taking $i=\alpha$, $j=\beta$ and $j=\alpha$, $i=\beta$. This yields

$$2g^{\alpha\alpha}\left(\frac{\partial}{\partial x_\alpha}\left\{\begin{array}{c}\alpha\,\beta\\ \beta\end{array}\right\}-\frac{\partial}{\partial x_\beta}\left\{\begin{array}{c}\alpha\,\alpha\\ \beta\end{array}\right\}\right)+2g^{\beta\beta}\left(\frac{\partial}{\partial x_\alpha}\left\{\begin{array}{c}\beta\,\alpha\\ \beta\end{array}\right\}-\frac{\partial}{\partial x_\alpha}\left\{\begin{array}{c}\beta\,\beta\\ \alpha\end{array}\right\}\right)=$$

$$g^{\alpha\alpha}\frac{\partial}{\partial x_\alpha}\left(g^{\beta\beta}\frac{\partial g_{\beta\beta}}{\partial x_\alpha}\right)+g^{\alpha\alpha}\frac{\partial}{\partial x_\beta}\left(g^{\beta\beta}\frac{\partial g_{\alpha\alpha}}{\partial x_\beta}\right)+g^{\beta\beta}\frac{\partial}{\partial x_\beta}\left(g^{\alpha\alpha}\frac{\partial g_{\alpha\alpha}}{\partial x_\beta}\right)+g^{\beta\beta}\frac{\partial}{\partial x_\alpha}\left(g^{\alpha\alpha}\frac{\partial g_{\beta\beta}}{\partial x_\alpha}\right).$$

For the second part of $2G$,

$$2G_2 = 2\sum_{ijk} g^{ii}\left(\left\{\begin{array}{c}i\,j\\ j\end{array}\right\}\left\{\begin{array}{c}k\,i\\ j\end{array}\right\}-\left\{\begin{array}{c}i\,i\\ k\end{array}\right\}\left\{\begin{array}{c}k\,j\\ j\end{array}\right\}\right)$$

is zero, if which $i=j=k$. In the set of terms that have two of the indices $i,j,k$ equal to each other and different from the third, we search out those having the like indices $\alpha$ and the other $\beta$, or the other way around. If $i=j=\alpha$ and $k=\beta$, we find zero, but, if $i=k=\alpha$, $j=\beta$ or $j=k=\alpha$, $i=\beta$, the results are, respectively

$$2g^{\alpha\alpha}\left(\left\{\begin{array}{c}\alpha\,\beta\\ \alpha\end{array}\right\}\left\{\begin{array}{c}\alpha\,\alpha\\ \beta\end{array}\right\}-\left\{\begin{array}{c}\alpha\,\alpha\\ \alpha\end{array}\right\}\left\{\begin{array}{c}\alpha\,\beta\\ \beta\end{array}\right\}\right) \text{ and } 2g^{\beta\beta}\left(\left\{\begin{array}{c}\beta\,\alpha\\ \alpha\end{array}\right\}\left\{\begin{array}{c}\alpha\,\beta\\ \alpha\end{array}\right\}-\left\{\begin{array}{c}\beta\,\beta\\ \alpha\end{array}\right\}\left\{\begin{array}{c}\alpha\,\alpha\\ \alpha\end{array}\right\}\right)$$

This is together equal to zero, because of (20). Of $G_2$ only those terms remain, therefore, which arise when we take all three, $i,j$ and $k$, different from each other. However, because it follows from (20), that a symbol of CHRISTOFFEL is zero for three different indices, the only term that remains in $2G_2$ is

$$-2\sum_{ijk} g^{ii}\left\{\begin{array}{c}i\,j\\ j\end{array}\right\}\left\{\begin{array}{c}k\,j\\ j\end{array}\right\}$$

We assign to the combination $\alpha\beta$ that part of this expression, that arises by taking $i=\alpha, j=\beta$ and $i=\beta, j=\alpha$. This yields

$$g^{\alpha\alpha}g^{\beta\beta}\left(g^{\gamma\gamma}\frac{\partial g_{\alpha\alpha}}{\partial x_\gamma}\frac{\partial g_{\beta\beta}}{\partial x_\gamma}+g^{\delta\delta}\frac{\partial g_{\alpha\alpha}}{\partial x_\delta}\frac{\partial g_{\beta\beta}}{\partial x_\delta}\right),$$

in which $\gamma$ and $\delta$ represent the two indices, that differ from $\alpha,\beta$ and from each other.

Herewith it is shown that $2G$ consists of six terms, each related to a combination of two indices. The term belonging to $\alpha\beta$ is

$$2G_{x_\alpha x_\beta} = g^{\alpha\alpha}\frac{\partial}{\partial x_\alpha}\left(g^{\beta\beta}\frac{\partial g_{\beta\beta}}{\partial x_\alpha}\right)+g^{\alpha\alpha}\frac{\partial}{\partial x_\beta}\left(g^{\beta\beta}\frac{\partial g_{\alpha\alpha}}{\partial x_\beta}\right)+g^{\beta\beta}\frac{\partial}{\partial x_\beta}\left(g^{\alpha\alpha}\frac{\partial g_{\alpha\alpha}}{\partial x_\beta}\right)+$$

$$+g^{\beta\beta}\frac{\partial}{\partial x_\alpha}\left(g^{\alpha\alpha}\frac{\partial g_{\beta\beta}}{\partial x_\alpha}\right)+g^{\alpha\alpha}g^{\beta\beta}\left(g^{\gamma\gamma}\frac{\partial g_{\alpha\alpha}}{\partial x_\gamma}\frac{\partial g_{\beta\beta}}{\partial x_\gamma}+g^{\delta\delta}\frac{\partial g_{\alpha\alpha}}{\partial x_\delta}\frac{\partial g_{\beta\beta}}{\partial x_\delta}\right)\quad\ldots(21).$$

18. If we take for the combination $x_\alpha x_\beta$ in succession $tr, t\vartheta, t\varphi, r\vartheta, r\varphi$ and $\vartheta\varphi$, we find from (21) without difficulties, because $g^{ii}=1/g_{tt}$ is,

$$2G_{tr} = \frac{1}{w^2}\frac{d}{dr}\left(-\frac{1}{u^2}\frac{d(w^2)}{dr}\right)-\frac{1}{w^2}\frac{d}{dr}\left(\frac{1}{w^2}\frac{d(w^2)}{dr}\right) = -\frac{4w''}{u^2w}+\frac{4u'w'}{u^3w},$$

$$2G_{t\vartheta} = 2G_{t\varphi} = -\frac{1}{w^2u^2v^2}\frac{d}{dr}\left(\frac{1}{w^2}\frac{d(w^2)}{dr}\right) = -\frac{4v'w'}{u^2vw},$$

$$2G_{r\vartheta} = 2G_{r\varphi} = -\frac{1}{u^2}\frac{d}{dr}\left(-\frac{1}{v^2}\frac{d(-v^2)}{dr}\right)-\frac{1}{v^2}\frac{d}{dr}\left(-\frac{1}{u^2}\frac{d(-v^2)}{dr}\right)=$$

$$-\frac{4v''}{u^2v} + \frac{4u'v'}{u^3v},$$

$$2G_{\vartheta\varphi} = -\frac{1}{v^2}\frac{d}{d\vartheta}\left(-\frac{1}{v^2\sin^2\vartheta}\frac{d(-v^2\sin^2\vartheta)}{d\vartheta}\right)-$$

$$-\frac{1}{v^2\sin^2\vartheta}\frac{d}{d\vartheta}\left(-\frac{1}{v^2}\frac{d(-v^2\sin^2\vartheta)}{d\vartheta}\right)-\frac{1}{u^2v^4\sin^2\vartheta}\frac{d(v^2)}{dr}\frac{d(v^2\sin^2\vartheta)}{dr} = \frac{4}{v^2}-\frac{4v'^2}{u^2v^2},$$

in which the accents signify differentiations with respect to $r$.

So, it becomes

$$2G = \frac{4}{v^2} - \frac{4v'^2}{u^2v^2} - \frac{8v'w'}{u^2vw} - \frac{8v''}{u^2v} - \frac{8v'v'}{u^3v} - \frac{4w''}{u^2w} + \frac{4u'w'}{u^3w} \quad \text{.........(22)}$$

Because

$$\sqrt{-g} = uv^2w\sin\vartheta$$

(19) becomes, after division by 2,

$$\delta\int_{t_1}^{t_2}dt\int_0^\pi \sin\vartheta\,d\vartheta\int_0^{2\pi}d\varphi\int_{r_1}^{r_2}\left(uw - \frac{wv'^2}{u} - \frac{2vv'w'}{u} - \frac{2vwv''}{u} +\right.$$

$$\left.+ \frac{2u'v'vw}{u^2} - \frac{v^2w''}{u} + \frac{v^2u'w'}{u^2}\right)dr = 0,$$

if we choose for the integration area the area, for which $t_1 \leq t \leq t_2$ and $r_1 \leq r \leq r_2$. The integrations over $\varphi$, $\vartheta$ and $t$ can be performed and after dividing by $4\pi(t_2 - t_1)$ we obtain

$$\delta \int_{r_1}^{r_2} \left\{ -\frac{d}{dr}\left(\frac{v^2 w'}{u}\right) - \frac{d}{dr}\left(\frac{2vwv'}{u}\right) + \frac{wv'^2 + 2vv'w'}{u} + uw \right\} dr = 0,$$

or, because $u, v, w, v', w'$ are not varied at the borders $r_1$ and $r_2$,

$$\delta \int_{r_1}^{r_2} \left\{ \frac{wv'^2 + 2vv'w'}{u} + uw \right\} dr = 0 \quad \ldots\ldots(23)$$

From this variation theorem follows immediately

$$\left.\begin{aligned}
V_1 &\equiv \frac{wv'^2 + 2vv'w'}{u^2} - w = 0, \\
V_2 &\equiv \frac{wv'' + v'w' + vw''}{u^2} - (vw' + wv')\frac{u'}{u^2} = 0, \\
V_3 &\equiv \frac{2vv'' + v'^2}{u} - u - 2vv'\frac{u'}{u^2} = 0
\end{aligned}\right\} \quad \ldots\ldots(24)$$

19. If we imagine in (18) a new variable $r_0$ substituted in place of $r$, so that

$$r_0 = f(r)$$

and (18) becomes

$$ds^2 = w_0^2 dt^2 - u_0^2 dr_0^2 - v_0^2 \left(d\vartheta^2 + \sin^2 \vartheta d\varphi^2\right),$$

in which

$$w_0 = w, \quad v_0 = v, \quad u_0 = u\lambda \quad \left(\lambda = \frac{dr}{dr_0}\right)$$

If accents at letters with the index 0 differentiations with respect to $r_0$, we have

$$v' = v'_0/\lambda, \ w' = w'_0/\lambda, \ u = u_0/\lambda, \ dr = \lambda dr_0$$

and so (23) becomes

$$\delta \int_{f(r_1)}^{f(r_2)} \left\{ \frac{w_0 v'^2_0 + 2v_0 v'_0 \, 'w_0\,'}{u_0} + u_0 w_0 \right\} dr_0 = 0,$$

which has completely the same form as (23), so that from here equations must follow with the same dependence on $r_0, u_0, v_0, w_0$ as (24) on $r, u, v, w$. Obviously, it is also possible without using (23) to verify, that a substitution $r_0 = f(r)$, $u = u_0 \frac{dr_0}{dr}$ leaves the equations (24) unchanged.

It is clear that the three equations (24) cannot possibly determine the functions $u, v,$ and $w$ completely and that they must be dependent on each other. Indeed we have

$$2v' V_2 + w' V_3 = u \frac{dV_1}{dr} \quad \dotfill (25)$$

Because, if one calls the integrand in (23) $L$, then is

$$V_1 = -\frac{\partial L}{\partial u},$$

$$2V_2' = \frac{d}{dr}\left(\frac{\partial L}{\partial v'}\right) - \frac{\partial L}{\partial v},$$

$$V_3 = \frac{d}{dr}\left(\frac{\partial L}{\partial w'}\right) - \frac{\partial L}{\partial w},$$

from where follows

$$u'V_1 + 2v'V_2 + w'V_3 = \frac{d}{dr}\left(v'\frac{\partial L}{\partial v'} + w'\frac{\partial L}{\partial w'}\right) - \frac{dL}{dr},$$

If one now calls the two parts that make up $L$, $P$ and $Q$, then we have

$$v'\frac{\partial L}{\partial v'} + w'\frac{\partial L}{\partial w'} = 2P$$

and so

$$u'V_1 + 2v'V_2 + w'V_3 = 2\frac{dP}{dr} - \frac{d(P+Q)}{dr} = \frac{d(P-Q)}{dr}$$

but, because $uV_1 = P - Q$, (25) follows.

20. If one now wants to solve for $u$, $v$ and $w$ from (24), the following method is available.  One chooses the independent variable $r$ in such a way, that one of the functions $u$, $v$ and $w$ becomes known.  Then one finds both others from two of the equations (24).  However, one can not ignore the equation $V_1 = 0$, because then, according to (25), $V_2 = 0$ combined with $V_3 = 0$ does not lead necessarily to $V_1 = 0$, but rather to

$$u\frac{dV_1}{dr} = 0 \ .$$

Therefore, if one chooses initially in (18) one of the functions $u$, $v$, $w$ and forms from the variation theorem the differential equation, actually two equations, these cannot be $V_2 = 0$ and $V_3 = 0$, in other words, one should not initially choose $u$, but $v$ or $w$.

In my original calculation of the field I had immediately assumed $u = 1$. I solved the equations $V_2 = 0$ and $V_3 = 0$ in such a way that $w$ remained finite at infinity and with that I satisfied not only $dV_1/dr = 0$, but also $V_1 = 0$. The observation by Prof. LORENTZ, that, by not setting $u$ equal to 1 *before* the variation, one can obtain a not completely unnecessary equation, has led me to the relationship explained here.

21. To solve the system (24) we pose now

$$v = r,$$

so that (18) becomes

$$ds^2 = w^2 dt^2 - u^2 dr^2 - r^2 \left( d\vartheta^2 + \sin^2 \vartheta d\varphi^2 \right) \quad \text{............(26)}$$

We keep the equations $V_1 = 0$ and $V_3 = 0$, so that, according to (25), also $V_2 = 0$ is satisfied. We obtain

$$\left. \begin{array}{l} V_1 = \dfrac{w + 2rw'}{u^2} - w = 0, \\[6pt] V_3 = \dfrac{1}{u} - u - 2r\dfrac{u'}{u^2} = 0, \end{array} \right\} \quad \text{...................(27)}$$

From the last equation follows

$$\frac{d}{dr}\left( \frac{r}{u^2} \right) = 1$$

and so

$$\frac{1}{u^2} = 1 - \frac{\alpha}{r}$$

in which $\alpha$ a constant is. If one substitutes this into the first equation (27), one obtains

$$2(r-\alpha)w' = \frac{\alpha}{r}w$$

or

$$(r-\alpha)\frac{d(w^2)}{dr} = \frac{\alpha}{r}w^2,$$

from where it follows

$$w^2 = 1 - \frac{\alpha}{r},$$

So we find

$$ds^2 = \left(1 - \frac{\alpha}{r}\right)dt^2 - \frac{dr^2}{1 - \frac{\alpha}{r}} - r^2\left(d\vartheta^2 + \sin^2\vartheta d\varphi^2\right) \quad \ldots\ldots(28)$$

22. Also K. SCHWARTZSCHILD has calculated this field (Sitzungsberichte der Kön. Preuss. Akad. der Wissenschafter, 1916, page 189). He poses

$$R^3 = r^3 + \alpha^3$$

and gives then the same expression for $ds^2$ in $R$, as I did in $r$. As a result $u^2$ remains finite for $r = \alpha$. This can obviously be reached in an infinite number of ways; SCHWARTZSCHILD's choice was determined by the wish, that $\sqrt{-g} = r^2 \sin^2\vartheta$ holds. Indeed, if one poses $R = f(r)$, then one finds after introducing the variable $r$

$$\sqrt{-g} = R^2 \frac{dR}{dr} \sin \vartheta$$

and in order for this to be equal to $r^2 \sin^2 \vartheta$, one must have

$$R^3 = r^3 + C .$$

For $R = \alpha$ to agree with $r = 0$, we must have $C = \alpha^3$ ;

It is clear, that introducing a variable, which is contained in (28) in a complicated manner, is not necessary. Just like one cannot allow negative values for $r$ in SCHWARTZSCHILD's formula (or in that of NEWTON), we shall exclude from (28) values for $r$ that are smaller than $\alpha$. Then we do not have to worry about the discontinuity at $r = \alpha$. Values for $r$, smaller than $\alpha$, are excluded for the reason that $ds^2$ becomes negative for values corresponding to $dr = d\vartheta = d\varphi = 0$, that is, for a mass point at rest.

Meanwhile we can move the discontinuity also to 0 without complicating (28), namely by replacing $r$ in the formula by $r + \alpha$. Then we get

$$ds^2 = \frac{dt^2}{1 + \frac{\alpha}{r}} + \left(1 + \frac{\alpha}{r}\right) dr^2 - (r+\alpha)^2 \left(d\vartheta^2 + \sin^2 \vartheta d\varphi^2\right) \quad \ldots\ldots(29)$$

Which equation one wishes to prefer, the one by SCHWARTZSCHILD, or (28) or (29), remains a question of taste. After all, one must realize that the *coordinate* $r$ does not represent the *measured distance*. One is free to choose a coordinate (provided that its values contain the whole area where the observation takes place only once), but one choice will be more useful than another.

However, we will pay attention to one coordinate system in particular. In the old theory one considers the velocity of light to be everywhere the same in all directions.

Also in the general theory of relativity one can in the case of one center choose the coordinates in such a way, that the velocity of light for all directions becomes the same, but it remains a function of the distance to the center. Because, if we substitute $r = f(\rho)$ into (28), we obtain

$$ds^2 = \left(1 - \frac{\alpha}{r}\right)dt^2 - \left(\frac{dr}{d\rho}\right)^2 d\rho^2 \bigg/ \left(1 - \frac{\alpha}{r}\right) - r^2\left(d\vartheta^2 + \sin^2\vartheta\, d\varphi^2\right),$$

and we reach our goal by posing

$$r^2/\rho^2 = \left(\frac{dr}{d\rho}\right)^2 \bigg/ \left(1 - \frac{\alpha}{r}\right),$$

Hereby is $\rho$ given as a function of $r$ except for a constant factor. We require

$$\lim_{r \to \infty} r/\rho = 1$$

and obtain then

$$r = \rho\left(1 + \frac{\alpha}{4\rho}\right)^2 \quad \text{..................(30)}$$

by which the expression for $ds^2$ transforms into

$$ds^2 = \left\{1 - \frac{\alpha}{\rho\left(1 + \frac{\alpha}{4\rho}\right)^2}\right\}dt^2 - \left(1 + \frac{\alpha}{4\rho}\right)^4 \left\{d\rho^2 + \rho^2\left(d\vartheta^2 + \sin^2\vartheta\, d\varphi^2\right)\right\} \quad \text{..(31)}$$

As one can see, the variable $\rho$ leads to a complicated expression for $ds^2$. For this reason, we will not use it in the following, except, a few times, to interpret results.

We still want to calculate the natural distance from a point, for which the $r$ of formula (28) is given, to the closest point on the sphere $r = \alpha$. This distance is, according to (28), if we overlook the factor $l$ (see point 4),

$$\delta = \int_\alpha^r \frac{dr}{\left(1-\frac{\alpha}{r}\right)^{\frac{1}{2}}} = r\sqrt{1-\frac{\alpha}{r}} + \alpha \log\left(\sqrt{1-\frac{\alpha}{r}} + \sqrt{\frac{\alpha}{r}}\right) \quad\quad (32)$$

If one expresses $\delta$ in $\rho$, one finds the simple formula

$$\delta = \rho - \frac{\alpha^2}{16\rho} + \tfrac{1}{2}\alpha \log \frac{4\rho}{\alpha} \quad\quad (33)$$

The natural distance between two points with the same $r$, measured along the circle through both points, is, just as in the Euclidian geometry, $r\psi$, if $\psi$ represents the angle between the radii to both points.

If $\delta$ is very large compared to $\alpha$, then the ratios $r : \rho : \delta$ approach 1. The differences $r - \delta$ and $\rho - \delta$ become larger than any value, but infinitely small compared to $r$, $\rho$ or $\delta$ itself, on a logarithmic scale.

23. We will see in 26, that, starting from formula (28), a mass point, which is outside the sphere $r = \alpha$, can never come inside that sphere. The question arises now, whether it is permitted to consider an atom (ignoring its charge) as a gravitational center, for which $\alpha$ represents the radius. This question, must, I think, be answered negatively, because the acceleration experienced by a mass point at a large distance (compared to $\alpha$), is according to 42

$$\ddot{r} = -\frac{\alpha c^2}{2r^2},$$

if $c$ represents the speed of light. If one sets this equal to the acceleration given by NEWTON's formula

$$\ddot{r} = -k\frac{m}{r^2},$$

than it follows

$$\alpha = -\frac{2km}{c^2}.$$

An order of estimate of this for a molecule of a single-atomic compound with molecular weight $M$ yields

$$\alpha = 2M \times 10^{-52} \text{ centimeter},$$

whereas the existing estimates for this are of the order of $10^{-9}$ centimeter. If one assumes, that only a part of the mass is measured by $\alpha$, assigning another part to the charges, than $\alpha$ becomes even smaller. It is obviously imaginable, that the sphere $r = \alpha$ the nucleus of the atom represents and that the electrons are outside of this sphere.

## § 2. **Calculation of a mass point in the field of a single center.**

24. To study the motion of one mass point in the field of a single center, we make use of formula (28). If the coordinates of the point are $r$, $\vartheta$ and $\varphi$, and if we represent differentiations with respect to time with points above the letters, than the quantity $L$ of 12, with $\lambda = t$, is

$$L = \left(1 - \frac{\alpha}{r} - \frac{\dot{r}^2}{1 - \frac{\alpha}{r}} - r^2 \dot{\vartheta}^2 - r^2 \left(\sin^2 \vartheta\right) \dot{\varphi}^2 \right)^{\frac{1}{2}}$$

The variation principle

$$\delta \int_{t_1}^{t_2} L \, dt = 0,$$

gives us then, among other things, the equation

$$\frac{d}{dt}\left(\frac{\partial L}{\partial \dot{\varphi}}\right) = 0,$$

which tells us, that

$$\frac{\partial L}{\partial \dot{\varphi}} = -\frac{r^2 \sin^2 \vartheta \, \dot{\varphi}}{L}$$

does not change and, therefore, once it is equal to zero, will keep that value. Because we can choose the coordinate system $\vartheta$ and $\varphi$ always in such a way, that $\dot{\varphi}$ for a certain value of $t$ equals zero, and since $\dot{\varphi}$ will keep than that value always, so that $\varphi$ does not change, the motion will occur in a flat plane. We will choose the pole of that polar coordinate system $\vartheta$, $\varphi$ in such a way, that this will become the plane $\vartheta = \pi/2$.

Then $L$ becomes

$$L = \left(1 - \frac{\alpha}{r} - \frac{\dot{r}^2}{1 - \frac{\alpha}{r}} - r^2 \dot{\varphi}^2\right)^{\frac{1}{2}}$$

The equations of motion now read

$$\frac{d}{dt}\left(\frac{\partial L}{\partial \dot{r}}\right) - \frac{\partial L}{\partial r} = 0, \quad \frac{d}{dt}\left(\frac{\partial L}{\partial \dot{\varphi}}\right) = 0 \quad \ldots\ldots(34)$$

By multiplying the first equation by $\dot{r}$, the second by $\dot{\varphi}$ and by adding the products, one obtains

$$\frac{d}{dt}\left(L - \dot{r}\frac{\partial L}{\partial \dot{r}} - \dot{\varphi}\frac{\partial L}{\partial \dot{\varphi}}\right) = 0$$

or

$$\frac{d}{dt}\left(\frac{1 - \frac{\alpha}{r}}{L}\right) = 0 \quad \ldots\ldots(35)$$

Instead of the two equations in (34) we can consider the system consisting of (35) and the second equation in (34). Both systems are equivalent, as long as $\dot{r}$ differs from 0; in case of the circular motion we will have to go back to (34).

We obtain now

$$\frac{1 - \frac{\alpha}{r}}{L} = \text{constant}, \quad \frac{r^2 \dot{\varphi}}{L} = \text{constant},$$

and, therefore, also

$$\frac{r^2\dot{\varphi}}{1-\frac{\alpha}{r}} = \text{constant}$$

This yields the equations

$$\frac{1}{1-\frac{\alpha}{r}} - \frac{\dot{r}^2}{\left(1-\frac{\alpha}{r}\right)^3} - \frac{r^2\dot{\varphi}^2}{\left(1-\frac{\alpha}{r}\right)^2} = A \quad \text{...............................(36)}$$

and

$$\frac{r^2\dot{\varphi}}{1-\frac{\alpha}{r}} = B \quad \text{...............................................................(37)}$$

in which $A$ and $B$ are constants.

The equations (36) and (37) are the ones, which can also teach us about non-circular motions.

25. *Circular motion.* Suppose first that $\dot{r}$ is constantly 0. The first of the two equations (34) teaches us, that

$$\frac{\partial L}{\partial r} = 0$$

is, from which follows

$$\dot{\varphi}^2 = \frac{\alpha}{2r^3} \quad \text{..............................................................(38)}$$

Substituting this and $\dot{r} = 0$ into the expression

$$\left(\frac{ds}{dt}\right)^2 = 1 - \frac{\alpha}{r} - \frac{\dot{r}^2}{1-\frac{\alpha}{r}} - r^2\dot{\varphi}^2,$$

yields

$$\left(\frac{ds}{dt}\right)^2 = 1 - \frac{3\alpha}{2r},$$

so that only when

$$r > \tfrac{3}{2}\alpha,$$

the motion occurs with a speed, which is smaller than the speed of light.

Here, as in cases, which we still must discuss, the equations themselves do not offer restrictions to the motion about the boundaries of speed. If one wants to express, that the motion happens with speeds smaller than that of light (a speed, which itself depends on $r$), than we must require that $L^2 > 0$ is. Once a point moves with a speed, larger or smaller than that of light, then the speed will remain during the complete motion larger or smaller than that of light. Because, as we saw, in the case of circular motion, the velocity $r\dot{\varphi}$ is constant and the speed of light, which only depends on $r$, also, because $r$ remains constant. And, in case of types of motion, of which the discussion will follow, because of (36)

$$L^2 = \left(1-\frac{\alpha}{r}\right)^2 A$$

and $L^2$ has, therefore, the same sign as that of $A$. For motions with a speed smaller than that of light, $A$ is positive; if the speed is larger than the speed of light, $A$ is negative.

26. We now return to the general case and take $\dot{r} \neq 0$. Then (36) and (37) determine the motion completely. We can take the constant $B$ always positive (or zero); this can always be ensured by reversing the positive direction of $\dot{\varphi}$. If we solve (37) for the differential $dt$ and substitute this in (36), then this equation transforms into

$$\frac{1}{1-\frac{\alpha}{r}} - \frac{B^2}{r^4\left(1-\frac{\alpha}{r}\right)}\left(\frac{dr}{d\varphi}\right)^2 - \frac{B^2}{r^2} = A \qquad\qquad\qquad\qquad (39)$$

Solving this for $d\varphi$ we find

$$d\varphi = \frac{B\,dr}{r^2\sqrt{1-\left(A+\frac{B^2}{r^2}\right)\left(1-\frac{\alpha}{r}\right)}}$$

We now introduce a new integration variable: $\quad x = \dfrac{\alpha}{r},$

by which we obtain

$$d\varphi = \frac{-dx}{\sqrt{x^3 - x^2 + \dfrac{A\alpha^2}{B^2}x + \dfrac{(1-A)\alpha^2}{B^2}}}$$

From here we see, that $x$, and so also $r$, an elliptic function of $\varphi$ is. If $x_1$, $x_2$ and $x_3$ are the roots of the equation

$$x^3 - x^2 + \frac{A\alpha^2}{B^2}x + \frac{(1-A)\alpha^2}{B^2} = 0,$$

then we have

$$x_1 + x_2 + x_3 = 1, \quad x_2 x_3 + x_3 x_1 + x_1 x_2 = \frac{A\alpha^2}{B^2}, \quad x_1 x_2 x_3 = \frac{(A-1)\alpha^2}{B^2} \quad \ldots(40)$$

and the constants $x_1, x_2, x_3$ (linked by the relations $x_1 + x_2 + x_3 = 1$) take over the role of $A$ and $B$. All three are real or one is real and the other two are complex but complementary.

If we now introduce a new variable $z$ and three new constants $e_1$, $e_2$ and $e_3$ by the equations

$$x = z + \tfrac{1}{3},$$
$$x_1 = e_1 + \tfrac{1}{3},$$
$$x_2 = e_2 + \tfrac{1}{3},$$
$$x_3 = e_3 + \tfrac{1}{3},$$

then we obtain

$$\left. \begin{array}{l} e_1 + e_2 + e_3 = 0, \quad e_2 e_3 + e_2 e_1 + e_1 e_3 = \dfrac{A\alpha^2}{B^2} - \tfrac{1}{3}, \\[2mm] \qquad e_1 e_2 e_3 = \dfrac{(2A-3)\alpha^2}{3B^2} + \tfrac{2}{27} \end{array} \right\} \quad \ldots(41)$$

and

$$d\varphi = \frac{dz}{\sqrt{(z-e_1)(z-e_2)(z-e_3)}} \quad \text{...........................(42)}$$

From here one sees, that one can express $z$ in $\varphi$ using the $\zeta$ --function, which corresponds to the roots $e_1$, $e_2$ and $e_3$. We satisfy (42) by posing

$$z = \zeta\left(\tfrac{1}{2}\varphi - C\right) \quad \text{.......................................(43)}$$

in which $C$ is an integration constant, which can be complex. From (43) follows

$$\frac{\alpha}{r} = \tfrac{1}{3} + \zeta\left(\tfrac{1}{2}\varphi + C\right) \quad \text{.......................................(44)}$$

This is the *orbit equation*. The motion is known completely, as soon as one of the two quantities $r$ and $\varphi$ is expressed in terms of the time $t$. Now from (37) follows

$$Bdt = \frac{r^2 d\varphi}{1 - \dfrac{\alpha}{r}} = \frac{\alpha^2 d\varphi}{x^2(1-x)} = \frac{\alpha^2 d\varphi}{\left(z+\tfrac{1}{3}\right)^2 \left(\tfrac{2}{3}-z\right)} \quad \text{..........................(45)}$$

Because of (42) this becomes

$$Bdt = \frac{\alpha^2 dz}{\left(z+\tfrac{1}{3}\right)^2 \left(\tfrac{2}{3}-z\right)\sqrt{(z-e_1)(z-e_2)(z-e_3)}}, \quad \text{...(46)}$$

while from (45) and (43) follows

$$Bdt = \frac{\alpha^2 dz}{\left\{\tfrac{1}{3}+\zeta\left(\tfrac{1}{2}\varphi+C\right)\right\}^2 \left\{\tfrac{2}{3}-\zeta\left(\tfrac{1}{2}\varphi+C\right)\right\}},$$

from where one sees, that $t$ and $\varphi$ can be expressed by $\zeta$'s and logarithms of $\sigma$'s.

We will, however, in the following not as a rule watch the progress of the motion with time, but only keep track of the orbit only, so that only (43) plays an important role. (46) will only serve for following the borders of the orbits. Because if the time, needed to reach a certain point in the orbit, is infinitely large, then such a point will never be reached nor crossed. From (46) follows, that $z$ never reaches $-\frac{1}{3}$ nor $+\frac{2}{3}$. This means that $r$ will never approach $\infty$ nor $\alpha$. We are used to not being able to reach infinity, but not getting to $\alpha$ means that reaching the sphere $r = \alpha$ is not possible. If we had, instead of using the $r$ from (28) we had used the one from (29), we could have said that the mass point will not reach the center. This result is completely different from what happens in NEWTON's theory; from this we find how different the motion is near the center (and also far away, if the speeds are large) compared to the classical theory. On the other hand, the deviations at medium speeds at large distances (compared to $\alpha$) are small, as we will see in more detail when discussing the motion of planets and comets in the gravitational field of the sun. One should keep in mind, however, that in reality, as soon as $r$ becomes smaller than a certain quantity $R$, the mass point probably will end up inside the matter, that makes up the spherical body generating the field, and that there $ds^2$ will be represented by another expression than we used until now.

     27. We now reach the discussion of the different cases for motion that can exist. The cases having $\dot{r}$ and $\dot{\varphi}$ always equal to zero, immediately distinguish themselves mechanically; the first case is that of the circular motion discussed in 25, the second is that of a mass point moving on a radius vector. From a mathematical point of view, the cases with two or three identical $e$ values distinguish themselves, because then we do not need elliptical functions for the integration. In these cases appears yet the peculiarity, that $z$ can only approach the multiple root, because according to (46) the time passed becomes infinite with decreasing distance.

When we now want to follow the motion using (42), (43) or (44) on the one hand and (46) on the other hand, we must realize, that $r$ lies always between $\infty$ and $\alpha$, so that $z$ lies between $-\frac{1}{3}$ and $\frac{2}{3}$, and that $(z-e_1)(z-e_2)(z-e_3)$ must always be positive. In other words, the constant $C$ in (43) and the range of values assumed by $\varphi$, must be chosen such that $z$ becomes real and lies between $-\frac{1}{3}$ and $\frac{2}{3}$.

As for the values of $e_1$, $e_2$ and $e_3$, all these three can be real (we assume then always $e_1 \geq e_2 \geq e_3$, $e_1$ is never negative, $e_2$ never positive) or two can be complex conjugated, while the third real is. In the first case $z$ must be larger than $e_1$ or must lie in between $e_2$ and $e_3$; in the second case $z$ must be greater than than the real root.

28. Now suppose first, that $z$ lies in between $e_2$ and $e_3$, while $e_2$ and $e_3$ lie within the interval $\left(-\frac{1}{3},\frac{2}{3}\right)$. Because $e_1 \geq e_2$, it follows from $e_1 + e_2 + e_3 = 0$, that then $e_2 \leq -\frac{1}{2}e_3$ must be, which at any rate is less than $\frac{1}{6}$. $r$ lies in between $\alpha/\left(\frac{1}{3}+e_3\right)$ and $\alpha/\left(\frac{1}{3}+e_2\right)$, this is between a value $r_1$, which can be unlimitedly large, but in any case larger is than $3\alpha$, and a value $r_2$, which is larger than $2\alpha$ and smaller than $r_1$. If we set, as is usual,

$$e_1 = \zeta\omega_1, \quad e_2 = \zeta\omega_2, \quad e_3 = \zeta\omega_3, \quad \omega_2 = \omega_1 + \omega_3$$

(so that $2\omega_1$ becomes the real period and $2\omega_3$ the purely imaginary period, in the case that the $e$ 's are real), then we see, that $\frac{1}{2}\varphi + C$ in (43) and (44) must have an imaginary part, which is congruent with $\omega_3$. If we propose, therefore, $C = \omega_3$, then $\frac{1}{2}\varphi$ runs through real values, actually through *all* of them. If $\varphi = 0$, then $z = e_3$, $r = r_1$; if $\varphi$ increase up to $2\omega_1$ then $z$ will increase up to $e_2$ and $r$ decreases to $r_2$; if $\varphi$ increases still further, then, when $\varphi$ has reached the value $4\omega_1$, $z$ and $r$ become again $e_3$ and $r_1$. The motion is, as far as $r$ is concerned, periodic, but because (in general) $4\omega_1 \neq 2\pi$, the orbit is not closed. This general motion should be called $(e_3, e_2)$, as $z$ oscillates between $e_3$ and $e_2$.

A special case, in which the $\zeta-\text{function}$ is degenerate, arises, if $e_2$ becomes equal to $e_1$. Then $z$ cannot reach the value $e_2$ and, therefore, $r$ cannot reach the value $r_2$. Because $e_2$ is positive, $r_2$ is smaller than $3\alpha$ $(\text{and} > 2\alpha)$. There is no periodicity anymore, but while $r$ will decrease from $r_1$ to $r_2$, $\varphi$ must increase to an infinitely large number, so that the orbit will wind itself as a spiral around the circle $r_2$; before the point reached the distance $r_1$ it came into a spiral that is symmetrically situated around the nearby circle $r_2$. This case is indicated as $(e_3, e_2 = e_1)$,

29. If one imagines in case of the motion $(e_3, e_2)$ that the root $e_3$ is shifted to $-\frac{1}{3}$, then we have $r_1 = \infty$. The orbit is extended all the way to infinity and the time needed to reach $r_1$, becomes according to (46) infinitely large. The value for $r_2$ is still bigger than $2\alpha$. We indicate this motion by $(e_3 = -\frac{1}{3}, e_2)$.

If in this motion, we have $e_2 = e_1 = \frac{1}{6}$, then $r_2$ reaches the value $2\alpha$ exactly, and the $\zeta-\text{function}$ becomes degenerate again. The motion is as follows: the mass point arrives from infinity and continues in a spiral having an infinite number of windings approaching the circle with radius $2\alpha$; or, the motion is in the opposite direction. This is the motion $\left(e_3 = -\frac{1}{3},\ e_2 = e_1\right)$.

30. If $e_3$ becomes even smaller, then $z$ will not reach the value $e_3$ any more; we obtain the motion $\left(-\frac{1}{3},\ e_2\right)$, at which we must have $-\frac{1}{3} < e_2 < \frac{2}{3}$ with $r$ varying between $\infty$ and $r_2$; now $r_2$ can be smaller than $2\alpha$. The point approaches from infinity, turns around the center and departs again towards infinity.

Also in this case we can have $e_2 = e_1$, so that the motion $\left(-\frac{1}{3},\ e_2 = e_1\right)$ arises. The $\zeta-\text{function}$ is degenerate and the orbit, arriving from infinity, winds around $r_2$.

31. If, at the motion $\left(-\frac{1}{3},\ e_2\right)$ the root $e_3$ becomes equal to or smaller than $-\frac{4}{3}$, than $e_2$ can become equal to $\frac{2}{3}$. The motion $\left(-\frac{1}{3},\ e_2 = \frac{2}{3}\right)$, which takes place in that case, is as follows (or in the opposite direction): the mass point arrives from infinity and approaches the circle with radius $\alpha$ more and more, whereby $\varphi$ increases from a certain finite value to $2\omega_1$. The motion will not proceed from there, however, because the point $\varphi = 2\omega_1$, $r = \alpha$ cannot be reached, because, according to (46) this takes an infinite amount of time.

If $e_3 = -\frac{4}{3}$ and $e_2 = e_1 = \frac{2}{3}$, exactly, then we have the degenerate case $\left(-\frac{1}{3},\ e_2 = e_1 = \frac{2}{3}\right)$, $2\omega_1$ becomes $\infty$ and so the orbit winds itself in the shape of a spiral around the circle $r = \alpha$.

If $e_3 < -\frac{4}{3}$, then $e_2$ can become also larger than $\frac{2}{3}$ and we obtain the motion $\left(-\frac{1}{3},\ \frac{2}{3}\right)$. The point approaches the circle $r = \alpha$ from infinity at a finite value for $\varphi$, smaller than $2\omega_1$. This value becomes $\infty$ (with $2\omega_1$), when $e_2$ becomes $e_1$ and the motion $\left(-\frac{1}{3}, \frac{2}{3}\right)_{e_2=e_1}$ arises; then the orbit winds itself in the shape of a spiral around the circle $r=\alpha$.

32. We now suppose that $z \geq e_1$ is. Then $e_1 \leq \frac{2}{3}$ must necessarily be true. If $e_1 = \frac{2}{3}$, then $z$ is equal to $e_1$ continuously, so we have $r=\alpha$. If we ignore this case, then we must have $e_1 < \frac{2}{3}$. We obtain the motion $\left(e_1, \frac{2}{3}\right)$, whereby $z$ varies between $e_1$ and $\frac{2}{3}$, $r$ between $\alpha/\left(\frac{1}{3}+e_1\right)$ and $\alpha$, noting that the value of $\alpha/\left(\frac{1}{3}+e_1\right)$ itself is between $\alpha$ and $3\alpha$). In (43) and (44) $C=0$ (or a multiple of $2\omega_3$) and $\varphi$ varies from a minimum value to a maximum value, where the minimum value lies between 0 and $2\omega_1$ and the maximum value between $2\omega_1$ and $4\omega_1$ minus this minimum value. During this motion $r$ climbs from $\alpha$ to $\alpha/\left(\frac{1}{3}+e_1\right)$ and then decreases again. The extreme parts of the orbit are completed infinitely slowly.

Three special cases are possible. First, $e_3 = e_2$ is possible; the motions keeps the same character. Let it be called $\left(e_1, \frac{2}{3}\right)_{e_2=e_1}$.

Second, $e_1 = e_2$ is possible. In this case the point leaves the circle $\alpha$ and continues in a spiral at the inside of the circle $r = \alpha/\left(\frac{1}{3}+e_1\right)$, getting closer and closer to this circle. This is the case $\left(e_2 = e_1, \frac{2}{3}\right)$.

Third, $e_3 = e_2 = e_1 = 0$ is possible. We then have the motion $\left(0, \frac{2}{3}\right)$, at which the point describes a spiral, leaving the circle $\alpha$ and approaching the circle $3\alpha$ from the inside.

33. We now come to the case that two $e$'s are complex conjugated; call these roots $e_1$ and $e_3$. $z$ must now be $\geq e_2$ and, therefore, we must have $e_2 < \frac{2}{3}$. If $e_2 > -\frac{1}{3}$, then we deal with case $\left(e_2, \frac{2}{3}\right)$, at which $z$ varies from $\frac{2}{3}$ up to $e_2$, and so $r$ varies from $\alpha$ to $r_2 = \alpha/\left(\frac{1}{3} + e_2\right)$; $r_2$ can have any values between $\alpha$ and $\infty$. In (43) and (44) if $C = 0$; $\varphi$ varies from a value $\varphi_0$ (situated between 0 and $2\omega_2$), for which holds

$$\zeta\left(\tfrac{1}{2}\varphi_0\right) = \tfrac{2}{3},$$

up to $4\omega_2 - \varphi_0$. The orbit leaves the circle $\alpha$ and returns to it.

If $e_2 = -\frac{1}{3}$, we obtain the motion $\left(e_2 = -\frac{1}{3}, \frac{2}{3}\right)$, at which the orbit extends to infinity; if $e_2 < -\frac{1}{3}$, then the orbit also extends itself to infinity and the motion $\left(-\frac{1}{3}, \frac{2}{3}\right)$ has arisen. In each of these cases degeneracy is possible, because $e_1$ becomes equal to $e_3$. If $e_1 = e_3 < e_2$, then the motions arise, which are discussed in 32, and at which now $e_2$ has taken over the role of $e_1$. If $e_1 = e_3 > e_2$, then arise, if one calls the two roots that are equal $e_2$ and $e_1$, and the other $e_3$, the cases $\left(e_3, e_2 = e_1\right)$, $\left(e_3 = -\frac{1}{3}, e_2 = e_1\right), \left(-\frac{1}{3}, e_2 = e_1\right), \left(-\frac{1}{3}, e_2 = e_1 = \frac{2}{3}\right), \left(-\frac{1}{3}, \frac{2}{3}\right)_{e_2 = e_1}$, which have already been discussed above.

If $e_1 = e_2 = e_3 = 0$, then motion $\left(0, \frac{2}{3}\right)$ arises again, which is discussed in 32.

34. We have now dealt with all the cases, that can occur except the case where $\dot{\varphi}$ is continuously equal to zero. That case we will discuss in 42 and 43. However, before continuing with the discussion of the consequences of the equations of motion, we must answer the question: which of the motions occur at a speed, that is smaller than that of light?

To answer this question easily and to understand the answer easily, we will utilize graphic representation. Let $e_3$ and $e_2$ be the Cartesian coordinates of a point in a flat plane; $e_3$ the abscissa and $e_2$ the ordinate. We start with the motions discussed in 28, 29, 30 and 31, for which $z$ lies between $e_3$ and $e_2$. Because $e_3$ must be negative, $e_2$ can never become larger than $-\frac{1}{2}e_3$ and $e_2$ greater than $e_2$, only that part of the plane can be considered, that is situated between the straight lines $e_2 = e_3$ and $e_2 = -\frac{1}{3}e_3$, and to the left of the axis $e_3 = 0$. If one now also draws the lines $e_3 = -\frac{4}{3}$, $e_2 = \frac{2}{3}$, $e_2 = -\frac{1}{3}$, then the part of the plane under consideration becomes partitioned in several parts; the different cases arise, because the point with coordinates $e_3$ and $e_2$ can be located in those different parts or on their boundaries. Because $e_2$ must be greater than $-\frac{1}{3}$ for the motions 28, 29, 30 and 31, then only the area EACG bordered by the lines $e_2 = -\frac{1}{3}e_3$, $e_2 = e_3$, $e_3 = -\frac{1}{3}$, can yield us the motions 28, 29, 30 and 31. I have indicated in the parts of the diagram and near some of the lines in it, the symbols of the corresponding motions.

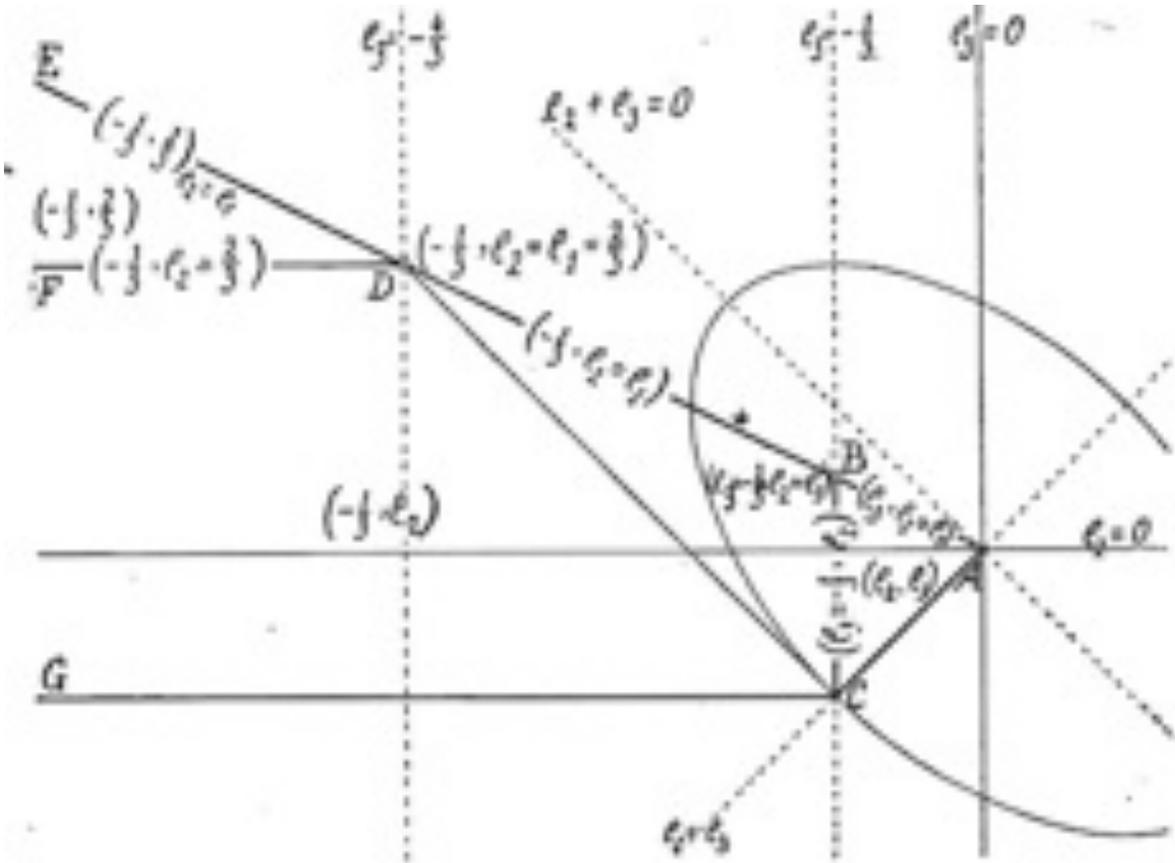

The question whether or not the speed at a motion exceeds that of light, comes down to the questions whether $A$ is positive or negative. (41) teaches us, that the motion occurs at a speed smaller than that of light, if

$$e_2 e_3 + e_3 e_1 + e_1 e_3 > -\tfrac{1}{3},$$

Replacing $e_1$ by $-e_2 - e_3$ transforms this condition to

$$e_3^2 + e_3 e_2 + e_2^2 < \tfrac{1}{3}.$$

The ellipse in our diagram is given by

$$e_3^2 + e_3 e_2 + e_2^2 = \tfrac{1}{3};$$

it passes through the point $e_3 = e_2 = -\tfrac{1}{3}$, which is one of the endpoints of the short axis. Only points inside the ellipse correspond to motions, for which the speeds are smaller than that of light. From there we see immediately, that the motions $(e_3, e_2)$, $(e_3, e_2 = e_1), (e_3 = -\frac{1}{3}, e_2), (e_3 = -\frac{1}{3}, e_2 = e_1)$ occur with speeds smaller that that of light. The motions $(-\frac{1}{3}, e_2), (-\frac{1}{3}, e_2 = e_1)$ can happen at larger and at smaller speed than that of light; the motions $(-\frac{1}{3}, \frac{2}{3}), (-\frac{1}{3}, \frac{2}{3})_{e_2=e_1}, (-\frac{1}{3}, e_2 = \frac{2}{3})$ and $(-\frac{1}{3}, e_2 = e_1 = \frac{2}{3})$ only occur at speeds larger than that of light.

35. It is clear, that a diagram, as we made constructed from the two quantities $e_2$ and $e_3$, can also be made using $e_1$ and $e_2$, or $e_1$ and $e_3$. Then

$$e_1^2 + e_1 e_2 + e_2^2 < \tfrac{1}{3} \text{ or } e_1^2 + e_1 e_3 + e_3^2 < \tfrac{1}{3}$$

becomes the condition to be satisfied, in order that the speed is smaller than the speed of light. We will, however, also in the cases of 32, keep using the $e_3$, $e_2$ diagram. The condition $e_1 < \frac{2}{3}$ becomes $e_2 + e_3 > -\frac{2}{3}$, which means, that the point $e_3$, $e_2$ must lie above the line $e_2 + e_3 = -\frac{2}{3}$. The area where the point $e_3$, $e_2$ can be located is, therefore, the triangle ACD, bordered by the lines $e_2 = e_3, e_2 + e_3 = -\frac{2}{3}, e_2 = -\frac{1}{2}e_3$.

Part of that triangle is inside the ellipse, another part is outside. The motion $(e_1, \frac{2}{3})$ can therefore occur at speeds larger than the speed of light, but also with speeds smaller than that. Of the special cases $(e_1, \frac{2}{3})_{e_2=e_3}, (e_2 = e_1, \frac{2}{3})$ and $(0, \frac{2}{3})$ the first and the third occur at speeds smaller than that of light, while the second can happen both at larger and at smaller speeds. The symbols of these cases have not been indicated in the diagram.

36. To also investigate the case, whereby $e_1$ and $e_3$ are complex, we set

$$e_1 = p(\cos\omega + i\sin\omega),$$
$$e_3 = p(\cos\omega - i\sin\omega),$$

and exclude the cases $e_1 = e_3$ ($\omega = 0$ or $\pi$), because, these, as we saw in 33, belong to situations discussed before. Because

$$e_2 e_3 + e_3 e_1 + e_1 e_2 = e_1 e_3 - e_2^2 = p^2 - e_2^2,$$

the condition, that $e_2 e_3 + e_3 e_1 + e_1 e_2 > -\tfrac{1}{3}$ becomes $p^2 > e_2^2 - \tfrac{1}{3}$.

Now $e_1 + e_2 + e_3 = 0$ or $2p\cos\omega + e_2 = 0$.

One can now choose $p$ always so large that $p^2 > e_2^2 - \tfrac{1}{3}$, and, by a suitable choice of $\omega$, will still continue to satisfy $2p\cos\omega + e_2 = 0$. All motions discussed in 33 can happen with a speed smaller than that of light; this is the case, among other situations, if $e_2^2 < \tfrac{1}{3}$.

## § 3. Special cases of the motion of one mass point in the field of one center.

37. In this section we want to consider more closely:

$1°.$ The motions, whereby the $\zeta-$function becomes degenerate, including a discussion of circular motion.

$2°.$ The motions on the vector radius.

$3°.$ The motions of planets and comets.

The starting point is the equations (36), (37), (39) and (44).

38. Differential equation (39), or, what is the same,

$$\left(\frac{dz}{d\varphi}\right)^2 = (z-e_1)(z-e_2)(z-e_3) \dots\dots\dots\dots\dots\dots\dots\dots\dots\dots\dots\dots(47)$$

can be integrated by elementary functions, if two of the $e$ 's are equal to each other, call this value $\lambda$, so that the third becomes $-2\lambda$. The equations then becomes

$$\left(\frac{dz}{d\varphi}\right)^2 = (z-\lambda)^2(z-2\lambda) \dots\dots\dots\dots\dots\dots\dots\dots\dots\dots(48)$$

We distinguish three cases, namely, $\lambda = 0, \ \lambda > 0, \lambda < 0.$ The starting point is the equations (36), (37), (39) and (44).

Suppose first $\lambda = 0,$ Then we have

$$\pm d\varphi = z^{-\frac{3}{2}}dz,$$

from which follows

$$\frac{\alpha}{r} = \tfrac{1}{3} + \frac{4}{\varphi^2} \dots\dots\dots\dots\dots\dots\dots\dots\dots\dots\dots\dots\dots\dots\dots(49)$$

in here the integration constant, which only determines the direction $\varphi = 0,$ has been given a certain value.

From (49) one see, that the motion occurs in a spiral, which comes out of the circle $r = \alpha$ and approaches more and more the circle $3\alpha.$ The case we are dealing with is not different from $(0, \tfrac{2}{3})$, already discussed in 32.

39. We suppose now that $\lambda > 0$ and assume

$$\lambda = a^2.$$

Then (48) becomes

$$\left(\frac{dz}{d\varphi}\right)^2 = \left(z-a^2\right)^2\left(z+2a^2\right).$$

from here follows

$$\pm d\varphi = \frac{2dy}{y^2-3a^2}.$$

There are now two main cases, namely $1^e\, z > a^2$, $2^e\, z < a^2$. If $z > a^2$, then this will remain true all the time, because according to (46) the double root cannot be reached: also $z < a^2$ continues to hold, once $z$ is smaller than $a^2$. If $z > a^2$, and, therefore, $y^2 > 3a^2$, then $y$ becomes

$$y = \pm\left(a\sqrt{3}\right)\coth^2\left(\tfrac{1}{2}a\varphi\sqrt{3}\right)$$

or

$$\frac{\alpha}{r} - \tfrac{1}{3} = z = -2a^2 + 3a^2\coth^2\left(\tfrac{1}{2}a\varphi\sqrt{3}\right) \quad\ldots\ldots\ldots\ldots\ldots\ldots(50)$$

in which again the integration constant has been determined. If, on the other hand, If $z < a^2$, and, therefore, $y^2 < 3a^2$, then $y$ becomes

$$y = \pm\left(a\sqrt{3}\right)\tanh\left(\tfrac{1}{2}a\varphi\sqrt{3}\right)$$

or

$$\frac{\alpha}{r} - \tfrac{1}{3} = z = 2a^2 + 3a^2\tanh^2\left(\tfrac{1}{2}a\varphi\sqrt{3}\right), \quad\ldots\ldots\ldots\ldots\ldots\ldots(51).$$

We see, that formula (50) relates to the second special case of 32, namely to the motion $\left(e_2 = e_1,\ \tfrac{2}{3}\right)$.

Formula (51) refers to the second special case of 32, namely to the motions $\left(e_{3},\ e_{2}=e_{1}\right)$, as discussed in 28, to $\left(e_{3}=-\tfrac{1}{3},\ e_{2}=e_{1}\right)$, discussed in 29, to $\left(-\tfrac{1}{3},\ e_{2}=e_{1}\right)$, discussed in 30, $\left(-\tfrac{1}{3},\ e_{2}=e_{1}=\tfrac{2}{3}\right)$, discussed in 31,

40. Third, we assume that $\lambda < 0$ is; take

$$\lambda = -a^{2}.$$

Then (48) transforms into

$$\left(\frac{dz}{d\varphi}\right)^{2} = \left(z+a^{2}\right)^{2}\left(z-2a^{2}\right),$$

from which we see, that $z > a^{2}$. So we deal with the motion of 32, and in particular with the first special case, the case $\left(e_{1},\ \tfrac{2}{3}\right)_{e_{3}=e_{2}}$. We set

$$z = y^{2} + 2a^{2}$$

and find

$$4\left(\frac{dy}{d\varphi}\right)^{2} = \left(y^{2} + 3a^{2}\right)^{2},$$

from which follows

$$\pm d\varphi = \frac{2dy}{y^{2} + 3a^{2}}.$$

We obtain

$$y = \pm\left(a\sqrt{3}\right)\tan\left(\tfrac{1}{2}a\varphi\sqrt{3}\right)$$

and, therefore,

$$\frac{\alpha}{r} - \tfrac{1}{3} = x = 2a^2 + 3a^2 \tan^2\left(\tfrac{1}{2}a\varphi\sqrt{3}\right),$$

41. We now come to answering the question: what is the place of the circular motions among the others?

To answer this question we imagine that in (47) a certain choice of the constants $e_1$, $e_2$ and $e_3$ has been made. Equation (47) has then a general solution, for which up to now we only had a discussion, and a singular solution, which must satisfy

$$\frac{dz}{d\varphi} = 0,$$

We obtain from here and from (47), by eliminating $\dfrac{dz}{d\varphi}$,

$$(z-e_1)(z-e_2)(z-e_3) = 0,$$

from which it is clear, that only

$$z = e_1, \ z = e_2, \ z = e_3$$

can be singular solutions; they are indeed, because they satisfy (47). We see from here, that the circular motions are singular solutions of the differential equation (47).

It is now, as we saw in 24, not yet certain, that these solutions also satisfy the equations of motion (34); for that not only (36) and (37) but also (38) must be satisfied. From these three equations follows, however,

$$A = \frac{1 - \dfrac{3\alpha}{2r}}{\left(1 - \dfrac{\alpha}{r}\right)^2}, \quad B^2 = \frac{\tfrac{1}{2}\alpha r}{\left(1 - \dfrac{\alpha}{r}\right)^2}$$

and from it follows easily, that

$$x^3 - x^2 + \frac{A\alpha^2}{B^2}x + \frac{(1-A)\alpha^2}{B^2}$$

which transforms into

$$\left(x - \frac{\alpha}{r}\right)^2 \left(x - 1 + \frac{2\alpha}{r}\right),$$

and so becomes

$$(z - e_1)(z - e_2)(z - e_3) = \left(z + \tfrac{1}{3} - \frac{\alpha}{r}\right)^2 \left(z - \tfrac{2}{3} + \frac{2\alpha}{r}\right).$$

The circular motions are, therefore, singular solutions of the differential equations for the degenerate motions, that is, of the equation (48). If $\lambda$ is positive, then is $\alpha/r > \tfrac{1}{3}$, so $r < 3\alpha$; if $\lambda$ is negative, then is $\alpha/r < \tfrac{1}{3}$, so $r > 3\alpha$ and for if $\lambda = 0$ is $r$ exactly equal to then is $3\alpha$. The circles with a radius smaller than $3\alpha$, wrap around the orbits of the motions discussed in 39, those with a radius bigger than $3\alpha$, envelop the orbits described in 40 and the circle $3\alpha$ wraps around the spirals of 38.

42. Now we come to the motions, that happen on the radius vector and for which, therefore, $\dot{\varphi} = 0$ holds continuously. We can start with the equations (36) and (37), for which the first transforms into

$$\frac{1}{1-\frac{\alpha}{r}} - \frac{\dot{r}^2}{\left(1-\frac{\alpha}{r}\right)^3} = A \qquad (52),$$

while the second requires that $B = 0$, in which case it will be satisfied by $\dot{\varphi} = 0$. As one can see easily from (41), two $e$'s are infinitely large and the third is $\frac{2}{3} - 1/A$.

Even though the integration of (52) can be performed without a problem and only elementary functions are involved, we prefer to read the properties of these motions directly from the differential equations without intervention of integrals. Solving (52) for $\dot{r}$ yields

$$\dot{r}^2 = \left(1-\frac{\alpha}{r}\right)^2\left(1 - A + A\frac{\alpha}{r}\right) \qquad (53)$$

and from here, by differentiating with respect to time,

$$\ddot{r} = \frac{\alpha}{r^2}\left(1-\frac{\alpha}{r}\right)\left(1 - \tfrac{3}{2}A + \tfrac{3}{2}A\frac{\alpha}{r}\right) \qquad (54)$$

If one eliminates $A$ from (53) and (54) or if one differentiates (52) with respect to time, one obtains

$$\ddot{r} = -\frac{\alpha}{2r^2}\left(1-\frac{\alpha}{r}\right) + \frac{3\alpha}{2r^2}\frac{\dot{r}^2}{1-\frac{\alpha}{r}},$$

an equation showing that the acceleration does not change when the direction of motion does not change; also from it follows, that at rest we have

$$\ddot{r} = -\frac{\alpha}{2r^2}\left(1-\frac{\alpha}{r}\right),$$

which is always towards the center, is zero both for $r = \infty$ and for $r = 0$, and has its biggest value for $r = \frac{3}{2}\alpha$.

From (53) follows, that

$$1 - A + A\frac{\alpha}{r}$$

can never be negative. If $A$ is smaller than 1 (we only consider positive values of $A$, to which refer to motions that have speeds smaller than that of light), then this expression is positive for each value of $r$. The mass point arrives then from infinity and is on its way to $r = \alpha$, where the motion according to the equation that follows from (53):

$$dt = \frac{dr}{\left(1-\frac{\alpha}{r}\right)\sqrt{1-A+A\frac{\alpha}{r}}}$$

becomes infinitely slow; the motion can obviously also occur in the opposite direction. The acceleration $\ddot{r}$ in this motion reaches the value zero only once (except for $r = \alpha$ and $r = \infty$), if $A > \frac{2}{3}$ is, namely for

$$r = \frac{3A\alpha}{3A-2},$$

as follows from (54). Between this value and $r = \alpha$ the acceleration is positive and reaches a maximum, between $r = 3A\alpha/(3A-2)$ and $r = \infty$ the acceleration is negative (attraction); that attraction also has a maximum. If, however, $A > 1$, then we must have, because of (53):

$$r < \frac{A\alpha}{A-1}.$$

The mass point moves now away from the center and turns around at $r = A\alpha/(A-1)$. The value $r = 3A\alpha/(3A-2)$, at which the acceleration becomes zero, is smaller than the value $r = A\alpha/(A-1)$, at which the motion reverses. During the ascend (at first repulsion occurs) is, therefore, the motion initially accelerated and this ceases at $r = 3A\alpha/(3A-2)$, after which the motion is still ascending, but decelerated; the deceleration first increases and *can* before the reversing of the motion reach its maximum, but can also stay in its increasing mode up to the point where it reverses; this happens at $r = A\alpha/(A-1)$ and the return motion is initially accelerating; all states of motion repeat themselves in reversed order, so that at $r = 3A\alpha/(3A-2)$ the motion becomes decelerated and becomes at the approach of the sphere $r = \alpha$ infinitely slow.

As we saw, $r$ can have all values, if $A < 1$ is; if $A = 1$, then the disappearing mass point will just not return. If $A < \frac{2}{3}$, then the acceleration will never reach zero; according to (54) there is then always repulsion and the speed is highest at infinite distance; there is repulsion during the entire motion.

43. One can ask oneself, what the influence has been from the choice of the variable $r$ on the results. If we execute a substitution

$$r = f(\rho),$$

will the results still be the same in terms of $\rho$, as they were first with respect to the variable $r$?

Let us consider more closely, which condition must be met by such a function $f(\rho)$. It must be such, that $f'(\rho)$ should exist everywhere and will never be negative; the first requirement is necessary in order to execute the substitution of the new variable into the the expression for $L$, the second is desired by the uniqueness in the correspondence of the variable $r$ and $\rho$ (taken strictly, one can also require, that $f(\rho)$. is always negative; but this can be dropped, if we agree always to choose that one from the pair, $\rho$ and $-\rho$, which makes $f'(\rho)$ not negative). In addition, we will assume that $f'(\rho)$ will never become equal to zero, remains finite, and has a continuous derivative $f''(\rho)$.

One can now keep using $r$ next to $\rho$ as an additional variable; from

$$\dot{r} = f'(\rho)\dot{\rho}$$

it is clear, that $\dot{\rho}$ becomes equal to zero at the same places where $\dot{r}$ becomes zero. Also, it follows immediately, that for $A < 1$ all values for $r$ and, therefore, also for $\rho$, are assumed (to be clear: all values for $\rho$, which belong to the domain for the variable $\rho$, that is, that system of values for $\rho$, that corresponds to all positive values of $r$ greater than $\alpha$), and that for $A > 1$ this is no longer the case.

Concerning the acceleration, we have

$$\ddot{r} = f'(\rho)\ddot{\rho} + f''(\rho)\dot{\rho}^2.$$

For $r = \alpha$ we have $\dot{r} = \ddot{r} = 0$ and, because $f'(\rho)$ does not reach zero, $\dot{\rho}$ becomes zero too; therefore, for $r = \alpha$ also $\ddot{\rho}$ becomes equal to zero. One can, according to (54), choose $r$ so close to $\alpha$, that $\ddot{r}$ is positive; the same is true for $\ddot{\rho}$, because one can choose $r$ also so close to $\alpha$, that $f''(\rho)\dot{\rho}^2$ vanishes compared to $\ddot{r}$. From there it follows, however, in the case $A > 1$, whereby the motion reverses once, the acceleration $\ddot{\rho}$ is first positive, then zero, and negative after that.

For $A < 1$, where the increasing point will not reverse, $\ddot{r}$ will reach zero once, as we saw, if $A > \frac{2}{3}$. One can now ask, if also $\ddot{\rho}$, just like $\ddot{r}$, will reach zero once, for a limited range of $A$ values. Because $\ddot{\rho}$ is initially positive, this will happen, if $\ddot{\rho}$ becomes negative for very large values of $r$. We now have

$$f'(\rho)\ddot{\rho} = \ddot{r} - f''(\rho)\dot{\rho}^2 = \ddot{r} - \frac{f''(\rho)}{f'(\rho)^2}\dot{r}^2;$$

this becomes, after substituting $\ddot{r}$ and $\dot{r}^2$ from (54) and (53):

$$f'(\rho)\ddot{\rho} = \frac{\alpha}{r^2}\left(1 - \frac{\alpha}{r}\right)\left(1 - \tfrac{3}{2}A + \tfrac{3}{2}A\frac{\alpha}{r}\right) - \frac{f''(\rho)}{f'(\rho)^2}\left(1 - \frac{\alpha}{r}\right)^2\left(1 - A + A\frac{\alpha}{r}\right)^2$$

or

$$f'(\rho)\ddot{\rho} = \left(1 - \frac{\alpha}{r}\right)\frac{1}{r^2}\left[\alpha\left(1 - \tfrac{3}{2}A + \tfrac{3}{2}A\frac{\alpha}{r}\right) - r^2\frac{f''(\rho)}{f'(\rho)^2}\left(1 - \frac{\alpha}{r}\right)\left(1 - A + A\frac{\alpha}{r}\right)\right].$$

For very large values of $r$ is the sign of $\ddot{\rho}$ equal to that of the expression between square brackets. If we now assume, that

$$\lim_{r \to \infty} r^2 \frac{f''(\rho)}{f'(\rho)^2} = B$$

exists, then will that expression for $r = \infty$ become equal to

$$\alpha\left(1 - \tfrac{3}{2}A\right) - B(1 - A)$$

and, therefore, will be negative, if

$$B > \tfrac{3}{2}\alpha \frac{\tfrac{2}{3} - A}{1 - A} \quad\ldots\ldots\ldots\ldots\ldots\ldots\ldots\ldots\ldots\ldots\ldots\ldots(55)$$

So, if $B$ is positive, then all motions with $1 > A \geq \tfrac{2}{3}$, have the property, that $\ddot{\rho}$ becomes zero once, even if $A$ becomes a bit smaller that $\tfrac{2}{3}$. For $B = 0$, $\ddot{\rho}$ will behave just like $\ddot{r}$, because if $A > \tfrac{2}{3}$, $\ddot{\rho}$ will become zero once. If $B$ is negative, then $A$ shall be greater than $\tfrac{2}{3}$ in any case; if one writes (55) in the form

$$\frac{1}{1 - A} > 3 - \frac{2B}{\alpha},$$

then one sees, that also now $A$ can approach 1 so closely, that $\ddot{\rho}$ once again becomes equal to zero,

For the variables $\rho$ and $\delta$ in 22 and 23, $B$ is equal to 0 and $\tfrac{1}{2}\alpha$, respectively. Phenomena take place, therefore, which are similar to those happening when using the variable $r$.

44. We still have to consider other motions for a mass point in the field. Namely, those that can be considered to represent the motions of planets and comets around the sun. As we will see, these are $(e_3, e_2)$, $(e_3 = -\tfrac{1}{3}, e_2)$ and $(-\tfrac{1}{3}, e_2)$.

To come to the planetary motions, we realize, that, because of the small speeds, $A$ in (36) differs from 1 with only a small amount, and actually so small, that it is proportional to the square of a component of the velocity. However, this is, as the theory of NEWTON teaches, of the order of $\alpha/r$. The quantity $B$ from (37) is of the order of $\sqrt{\alpha r}$. We set, therefore,

$$A = 1 + \frac{\mu\alpha}{\lambda^2}, \quad B = \sqrt{\alpha}/\lambda,$$

in which $\lambda$ and $\mu$ are two constants, which take the place of $A$ and $B$; the first has the same order of magnitude as $1/\sqrt{r}$, the second of the same order as $1/r^2$. We can make sure that $\lambda$ becomes positive, by choosing the positive rotation direction of $\varphi$ such, that $\dot{\varphi} > 0$. Now equation (39) becomes

$$\frac{\lambda^2}{r-\alpha} - \frac{\left(\dfrac{dr}{d\varphi}\right)^2}{r^3(r-\alpha)} - \frac{1}{r^2} = \mu,$$

which for $\alpha = 0$ transforms into the equation

$$\frac{\lambda^2}{r} - \frac{1}{r^4}\left(\frac{dr}{d\varphi}\right)^2 - \frac{1}{r^2} = \mu, \quad\quad\quad\quad\quad\quad\quad\quad\quad\quad (56)$$

which appears in the theory of NEWTON and there an ellipse, parabola or hyperbola represents, according to $\mu$ being positive, zero or negative. From here it follows that in NEWTON's theory we have $\mu < \tfrac{1}{4}\lambda^4$.

After introducing the constants $\lambda$ and $\mu$ into the equations (40), these become

$$x_1 + x_2 + x_3 = 1, \quad x_2 x_3 + x_3 x_1 + x_1 x_2 = \alpha\left(\lambda^2 + \mu\alpha\right), \quad x_1 x_2 x_3 = \mu\alpha^2.$$

We see, that the roots $x_1$, $x_2$, $x_3$ are close to 1,0,0, the roots $e_1$, $e_2$, $e_3$ are, therefore, close to $\frac{2}{3}, -\frac{1}{3}, -\frac{1}{3}$. The quantity $\alpha(\lambda^2 + \mu\alpha)$ is positive.

We now compare the motion, corresponding with a certain set of values for $\lambda$ and $\mu$, with that motion from NEWTON's theory, for which $\lambda$ and $\mu$ have the same values; we obtain, as we saw, the last from the first by setting $\alpha = 0$. We will set, therefore, $\mu < \lambda^4$ even if $\alpha$ is not zero. It easily becomes evident, that $x_1$ is a little bit smaller than 1, actually about $\alpha\lambda^2$ smaller; $x_2$ and $x_3$ are of the order $\alpha$, and positive when $\mu$ is positive, but with a different sign for negative $\mu$. We set, therefore,

$$\left.\begin{array}{l} x_1 = 1 - 2\alpha m, \\ x_2 = \alpha(m+n), \\ x_3 = \alpha(m-n), \end{array}\right\} \quad \text{...............(57)}$$

Now is, as it should be, $x_1 + x_2 + x_3 = 1$; for $n < m$, one obtains the motion $(e_3, e_2)$, for $n = m$, then $(e_3 = -\frac{1}{3}, e_2)$, for $n > m$, then $(-\frac{1}{3}, e_2)$, which correspond, respectively with the elliptic, parabolic and hyperbolic motion in NEWTON's theory. The constants $n$ and $m$ take the place of the constants $\lambda$ and $\mu$.

As a result of (57) we obtain

$$\left.\begin{array}{l} e_1 = \frac{2}{3} - 2\alpha m, \\ e_2 = -\frac{1}{3} + \alpha(m+n), \\ e_3 = -\frac{1}{3} + \alpha(m-n), \end{array}\right\} \quad \text{...............(58)}$$

The constant $C$ in the formulas (43) and (44) is now $\omega_3$ and (4) becomes

$$\frac{\alpha}{r} = \tfrac{1}{3} + \zeta\left(\tfrac{1}{2}\varphi + \omega_3\right).$$

We can write this equation using the well-known formula

$$\zeta\left(\tfrac{1}{2}\varphi + \omega_3\right) = e_3 + \frac{(e_1 - e_3)(e_2 - e_3)}{\zeta\tfrac{1}{2}\varphi - e_3}$$

in the form

$$\frac{\alpha}{r} = \tfrac{1}{3} + e_3 + \frac{(e_1 - e_3)(e_2 - e_3)}{\zeta\tfrac{1}{2}\varphi - e_3},$$

and this, because of (58), becomes

$$\frac{1}{r} = m - n + \frac{2n(e_1 - e_3)}{\zeta\tfrac{1}{2}\varphi - e_3} \quad \text{...(59)}$$

One can also express $1/r$ in terms of $\varphi$ using JACOBI's function sn; one only needs to use the well-known formula

$$\mathrm{sn}^2\sqrt{e_1 - e_3} = \frac{e_1 - e_3}{\zeta u - e_3} \quad \left(k^2 = \frac{e_2 - e_3}{e_1 - e_3}\right)$$

to find

$$\frac{1}{r} = m - n + 2n\ \mathrm{sn}^2 \tfrac{1}{2}\varphi\sqrt{e_1 - e_3} \quad \text{...(60)}$$

For $\alpha = 0$ one obtains $e_1 - e_3 = 1$ and $k^2 = 0$; and (60) transforms into

$$\frac{1}{r} = m - n + 2n\ \sin^2 \tfrac{1}{2}\varphi = m - n\cos\varphi$$

and this is the orbit equation in NEWTON's theory.

45. Now we will consider the case of planetary motion more closely; $n$ is now positive. Equation (59) teaches, that $r$ will recover its former value, as soon as $\varphi$ has increased by $4\omega_1$. Because the $\zeta-\text{function}$ is now degenerate, we can set

$$\omega_1 = \frac{\pi}{\sqrt{(e_1-e_3)} + \sqrt{(e_1-e_2)}} \quad \text{.....................(61)}$$

and expand the roots in increasing powers of $\alpha$. We find then

$$4\omega_1 = 2\pi\left(1 + \tfrac{3}{2}\alpha m\right) = 2\pi + 3\alpha m\pi \quad \text{..................(62)}$$

If we had carried the approximation of $\omega_1$ further than done in (61), we would have found only in the higher powers of $\alpha$ other coefficients. From (62) we see, that at every revolution (that is, when $r$ has obtained its previous value again) $\varphi$ has increased with $2\pi + 3\alpha m\pi$. The perihelion motion is therefore $3\alpha m\pi$.

From (59) follows, that $m-n$ is the minimum and $m+n$ the maximum value of $1/r$. Therefore, $m$ is the inverse of the orbit parameter $p$ and $n/m$ represents the eccentricity $e$. So

$$e = \frac{n}{m} \; , \; p = \frac{1}{m}.$$

Therefore, the perihelion motion becomes $3\alpha\pi/p$, which is also what EINSTEIN finds.

46. We also want to calculate the revolution time. From formula (37) follows

$$Bdt = \frac{r^2 d\varphi}{1 - \dfrac{\alpha}{r}},$$

which becomes the corresponding formula in NEWTON's theory for $\alpha = 0$. We expand the numerator and cut off the expansion after the $\alpha/r$ term. We then obtain

$$Bdt = r^2\left(1+\frac{\alpha}{r}\right)d\varphi = r^2 d\varphi + \alpha r d\varphi \quad \text{.........................(63)}$$

If we set

$$\sin\psi = \operatorname{sn}\tfrac{1}{2}\varphi\sqrt{e_1 - e_3},$$

we find by differentiation

$$\tfrac{1}{2}\sqrt{e_1 - e_3}\,d\varphi = \frac{d\psi}{\sqrt{1 - k^2 \sin^2\psi}}$$

As (60) now transforms into

$$\frac{1}{r} = m - n + 2n\,\sin^2\psi,$$

we obtain from (63)

$$\tfrac{1}{2}B\sqrt{e_1 - e_3}\,dt = \frac{d\psi}{\left(m - n + 2n\sin^2\psi\right)^2 \sqrt{1 - k^2 \sin^2\psi}} +$$

$$+ \frac{\alpha d\psi}{\left(m - n + 2n\sin^2\psi\right)\sqrt{1 - k^2 \sin^2\psi}} \quad \text{........(64)}$$

Now we have

$$B^2 = \alpha/\lambda^2 = \alpha^2/\left(x_2 x_3 + x_3 x_1 + x_1 x_2 - x_1 x_2 x_3\right)$$

and, therefore according to (57)

$$B^2 = \alpha / \left[ 2m(1 - 2\alpha m + \alpha^2 m^2 - \alpha^2 n^2) \right],$$

that is, with sufficient accuracy

$$B^2 = \frac{\alpha}{2m}(1 + 2\alpha m),$$

Furthermore

$$e_1 - e_3 = 1 - 3\alpha m + \alpha n,$$

and

$$B^2(e_1 - e_3) = \frac{\alpha}{2m}(1 - \alpha m + \alpha n)$$

so that

$$B\sqrt{e_1 - e_3} = \sqrt{\frac{\alpha}{2m}}\left[1 - \tfrac{1}{2}\alpha(m-n)\right]$$

As a result, (64) becomes

$$\tfrac{1}{2}\sqrt{\frac{\alpha}{2m}}\left[1 - \tfrac{1}{2}\alpha(m-n)\right]dt = \frac{d\psi}{(m - n + 2n\sin^2\psi)^2 \sqrt{1 - k^2 \sin^2\psi}} +$$

$$+ \frac{\alpha d\psi}{(m - n + 2n\sin^2\psi)\sqrt{1 - k^2 \sin^2\psi}}$$

Because $k^2$ is of the order $\alpha$, we expand the square root in the numerator and ignore $k^4$ and higher powers of $k^2$; in the second fraction we can set $k^2 = 0$. If we set in the result $k^2 = 2\alpha n$, then we obtain

$$\tfrac{1}{2}\sqrt{\frac{\alpha}{2m}}\left[1-\tfrac{1}{2}\alpha(m-n)\right]dt = \frac{(1+\alpha n\sin^2\psi)d\psi}{(m-n+2n\sin^2\psi)^2} + \frac{\alpha d\psi}{m-n+2n\sin^2\psi} =$$

$$= \frac{(1-\tfrac{1}{2}\alpha(m-n))d\psi}{(m-n+2n\sin^2\psi)^2} + \frac{\tfrac{3}{2}\alpha d\psi}{m-n+2n\sin^2\psi}.$$

If we divide this equation by $1-\tfrac{1}{2}\alpha(m-n)$, we obtain

$$\tfrac{1}{2}\sqrt{\frac{\alpha}{2m}}dt = \frac{d\psi}{(m-n+2n\sin^2\psi)^2} + \frac{\tfrac{3}{2}\alpha d\psi}{m-n+2n\sin^2\psi}.$$

If we call $T$ the time in which $r$ is periodic, so in which $\varphi$ increases by $4\omega_1$ and $\psi$ by $\pi$, then we find

$$\tfrac{1}{2}\sqrt{\frac{\alpha}{2m}}T = \int_0^\pi \frac{d\psi}{(m-n+2n\sin^2\psi)^2} + \tfrac{3}{2}\alpha\int_0^\pi \frac{d\psi}{m-n+2n\sin^2\psi} =$$

$$\frac{\pi m}{(m^2-n^2)^{\frac{3}{2}}} + \frac{\tfrac{3}{2}\alpha\pi}{(m^2-n^2)^{\frac{1}{2}}}$$

or

$$\frac{\sqrt{\alpha}}{2\pi\sqrt{2}}T = \left(\sqrt{\frac{m}{m^2-n^2}}\right)^3 + \tfrac{3}{2}\alpha\sqrt{\frac{m}{m^2-n^2}} = a^{\frac{3}{2}} + \tfrac{3}{2}\alpha a^{\frac{1}{2}},$$

if $a$ represents half the size of the large axis. We can also write this in the form

$$\frac{\sqrt{\alpha}}{2\pi\sqrt{2}}T = (a+\alpha)^{\frac{3}{2}},$$

so that the third law of KEPLER obtains the simple form:

$$\frac{(a+\alpha)^3}{T^2} = \frac{\alpha}{8\pi^2} \quad \text{.................................................(65)}$$

One can also inquire about the time needed for $\varphi$ to increase by $2\pi$. It depends on the location, where one lets the planet start and is maximal for the perihelion, minimal for the aphelion. If after a time $NT$ the perihelion (or any other predetermined point of the orbit) has rotated over an angle $(N+1)2\pi$, then, according to (45), we

must have
$$2\pi = N\frac{3\alpha\pi}{p},$$

and, therefore,
$$N = \frac{2p}{3\alpha},$$

from where for the angle $(N+1)2\pi$ follows

$$2\pi\left(1+\frac{2p}{3\alpha}\right);$$

this angle is covered in the time

$$NT = \frac{2p}{3\alpha}T,$$

so that one can consider as an average revolution time the quantity

$$T_1 = \frac{2p}{3\alpha}T \bigg/ \left(1+\frac{2p}{3\alpha}\right) = T\bigg/\left(1+\frac{3\alpha}{2p}\right),$$

for which one from (65) easily the following relation finds

$$\left(a - \frac{\alpha e^2}{1-e^2}\right)^3 \bigg/ T_1^2 = \frac{\alpha}{8\pi^2},$$

which deviates less from KEPLER's law than (65), because $e$ is small for the orbit of a planet; for a circular orbit this law does not deviate at all from that of KEPLER.

on characteristics of the equations.

---

# CHAPTER III.

---

## THE FIELD OF SLOWLY MOVING MASSES

### § 1. **Introduction.**

47. In the previous chapter we calculated the field of one single motionless center, that possess spherical symmetry, and we studied the movement of one mass point in that field. Going from this simple case to more complicated situations, one could ask about the field of a moving center or of a number of moving center with known state of movement; and, on the other hand one could, for a given field, ask about the movements of a body for which the size cannot be ignored. The exact solution of these problems is not easy and it is, therefore, advisable, at least for the time being, to search for an approximate solution; these can be satisfactory, as far as applications are concerned, if the accuracy of the approximation is such, that the influence of the neglected terms cannot be observed. In this chapter, we will concern ourselves with the search of such an approximate solution for the equations of gravity.

48. Let us call the values for $g_{ij}$ and $g^{ij}$ in the absence of masses causing gravity $\alpha_{ij}$ and $\alpha^{ij}$, respectively, so that we have

$$\alpha_{ij} = 0 \text{ for } i \neq j, \ \alpha_{11} = \alpha_{22} = \alpha_{33} = -1, \ \alpha_{44} = 1, \ \alpha^{ij} = \alpha_{ij}$$

If we ignore in the equations of gravity the squares and products (and higher powers) of the differences between the $g$'s and the $\alpha$'s, then the first terms in these equations will become linear in those differences, and because the second term, if it is not zero, depends on a factor $\kappa$ (expressing these differences), then also in the second term the factor $\kappa$ will be present. We can then expand the $g$'s in powers of $\kappa$ and determine the terms by these expansions. If, however, the second term in each equation of gravity is zero, then the linear equations for the difference between the $g$'s and the $\alpha$'s, teach us how the force of gravity propagates. A paper on this topic by EINSTEIN has appeared recently, entitled "Näherungsweise Integration der Feldgleichungen der Gravitation" (Sitzungsberichte der Kön. Preuss. Akad. d. Wiss.1916, page 688), from which it appears that gravity propagates at the speed of light.

49. We shall limit ourselves to the approximation of the field of masses, for which the speeds are small compared to the speed of light, that is the speed, at which the actions of gravity propagate. This is the case, among other things, for planets.

If we have in a field of gravity a moving mass point, then we can expect, that the squares of the velocity components of the point will be of the same order of magnitude as the terms of the first order in the $g$'s, those are the terms containing the factor $\kappa$; we saw this also in the second chapter, where the square of the speed of the order $\alpha/r$ was.

Now we want to calculate the gravitation field so accurately, that we obtain also the second order terms from the equations of motion. According to 12, one can obtain those equations of motion by setting the first variation of

$$\int_{t_1}^{t_2} L\,dt = \int_{t_1}^{t_2} \left( \sum_{ij} g_{ij} \dot{x}_i \dot{x}_j \right)^{\frac{1}{2}} dt$$

equal to zero. One can now see easily that for the quantity $L$ the terms of the second order must be known. If we indicate an index assuming only 1, 2 and 3, by placing the index under the summation sign between parentheses, then we obtain

$$L^2 = \sum_{(j)} g_{ij} \dot{x}_i \dot{x}_j + 2 \sum_{(j)} g_{4j} \dot{x}_j + g_{44},$$

from which we see, realizing that $\dot{x}_i$ and $\dot{x}_i$ are of the order $\frac{1}{2}$, that $g_{44}$ up to terms in the second order, $g_{4j}$ up to terms of the order $1\frac{1}{2}$, $g_{ij}\,(i \neq 4, j \neq 4)$ up to terms in the first order must be calculated.

    50. Now we imagine that the field generating masses are in motion and under each others influence are in such a way, that their speeds are also of the order $\frac{1}{2}$. The field of the first order is, as we will see in 55, independent of the state of motion of the field generating bodies and equal to the sum of the first order fields, that are generated by each of the bodies separately. One can now make a mental picture before-hand of the influence, that the simultaneous presence of the bodies and their motions will have on the field. We saw in the second chapter, that in the case of one spherical center at rest $g_{14} = g_{24} = g_{34} = 0$. Let us first think about bodies at rest; due to various reasons, $g_{ij}\,(i \neq 4, j \neq 4)$ and $g_{44}$ will contain terms of the second order. In the first place, they arise, as in the case of a single center, because the equations, from which they are found after calculating terms of first order, contain squares and products of first order terms, as a result of the non-linearity of the first terms in the equations of gravity.

Now, because the first order terms for multiple centers are the sum of terms from each center separately, second order terms arise also from products of first order terms. This is the reason for the appearance of the third term in (99). A second reason for having second order terms is due to the potential energy of each center in the field of the other centers. This changes the value of $T_4^4$ with an amount of the first order and, as a result, causes in the $g$'s a term of the second order; this is the origin of the sixth term in (99). If we now also realize that the masses are moving resulting in a kinetic energy of the first order, causing first order changes in $T_4^4$ inducing second order terms in the $g$'s. This causes the fifth term in (99).

The motion of field generating masses has, however, still another effect. Due to the finite propagation speed of the gravitation it is necessary to calculate the field in a point at a certain moment by taking into account the positions of the field generating bodies at the moment that they emitted the field. These are positions, which deviate from the instantaneous positions with quantities of the order $\frac{1}{2}$, because the speeds are of the order $\frac{1}{2}$. These so called effective positions can also be calculated accurately to the first order, which will be necessary to describe the $g$'s (in particular n the $g_{44}$) correctly to the second order. A closer calculation teaches, that quantities of the order $\frac{1}{2}$ in $g_{44}$ (and $g_{ij}$ for $i \neq 4$ and $j \neq 4$, but those are not relevant here) are not present, but there are only terms of the second order, caused by the acceleration and the squares of velocity components. This is one of the reasons for the presence of the fourth term in (99); we do not, however, obtain this term from the calculation of the effective positions. The quantities $g_{14}$, $g_{24}$ and $g_{34}$ get a term of the order $1\frac{1}{2}$, which can be understood easiest by realizing the gravitational action of the quantities $T_{i4}$, which are of the order $\frac{1}{2}$ for $i \neq 4$. As a result, $g_{i4}$ in $ds^2$ becomes different from zero (compare formula (100)); these quantities also play a part in the origin of the fourth term in (99), due to their presence in the first part of the equations of gravity.

51. Before we start with the direct calculation of the field generated by a number of moving masses, we want, in order to clarify what is said in 50, to calculate the field in a simple case. We imagine a center moving on a straight line at constant velocity, $v$, along the $x$-axis. It is spherically symmetric in a coordinate system in which it is at rest; it will then be flattened in the coordinate system in which it moves.

In the second chapter we found formula (31) when we used the variable $\rho$. This equation can also be written as

$$ds^2 = \left(\frac{1-\frac{\alpha}{4\rho}}{1+\frac{\alpha}{4\rho}}\right)^2 d\tau^2 - \left(1+\frac{\alpha}{4\rho}\right)^4 \left(d\xi^2 + d\eta^2 + d\zeta^2\right),$$

if we call the time $\tau$ and introduce the Cartesian coordinate system $\xi, \eta, \zeta$ with

$$\rho^2 = \xi^2 + \eta^2 + \zeta^2.$$

After expanding this in powers of $\alpha$ and cutting the expansion of $d\tau^2$ after the $\alpha^2$ term and that of the coefficient of $d\xi^2 + d\eta^2 + d\zeta^2$ after the $\alpha$ term, we find

$$ds^2 = \left(1 - \frac{\alpha}{\rho} + \frac{\alpha^2}{2\rho^2}\right) d\tau^2 - \left(1 + \frac{\alpha}{\rho}\right)\left(d\xi^2 + d\eta^2 + d\zeta^2\right) \quad \text{......(66)}$$

This formula refers to a center, that is at rest in the point $\xi = \eta = \zeta = 0$. Now we consider a transformation used in the special theory of relativity. Suppose

$$\xi = ax - bt, \ \eta = y, \ \zeta = z, \ \tau = at - bx \quad \text{..............(67)}$$

in which is

$$a = \left(1 - v^2\right)^{-\frac{1}{2}}, \ b = v\left(1 - v^2\right)^{-\frac{1}{2}}$$

and $v$ represents; one has then, as required,

$$a^2 - b^2 = 1.$$

From (67) follows

$$x = vt + \xi\sqrt{1 - v^2},$$

so that a point, at rest in the coordinate system $\xi, \eta, \zeta, \tau$, is moving in the $x, y, z, t,$ system at constant velocity along the x-axis. The transformation (67) is such, that we have

$$t^2 - \left(x^2 + y^2 + z^2\right) = \tau^2 - \left(\xi^2 + \eta^2 + \xi^2\right);$$

the values of the $g$'s at infinity remain the same as before.

Substituting from (67) the following values

$$d\xi = adx - bdt, d\xi = adx - bdt, \ d\eta = dy, \ d\zeta = dz, \ d\tau = adt - bdx$$

into (66), then (66) becomes

$$ds^2 = \left(1 - \frac{\alpha}{\rho} + \frac{\alpha^2}{2\rho^2}\right)(adt - bdx)^2 - \left(1 + \frac{\alpha}{\rho}\right)\left((adx - bdt)^2 + dy^2 + dz^2\right)$$

or

$$ds^2 = \left\{\left(1 - \frac{\alpha}{\rho} + \frac{\alpha^2}{2\rho^2}\right)a^2 - \left(1 + \frac{\alpha}{\rho}\right)b^2\right\}dt^2 + 2ab\left(\frac{2\alpha}{\rho} - \frac{\alpha^2}{2\rho^2}\right)dxdt -$$

$$-\left\{\left(1 + \frac{\alpha}{\rho}\right)a^2 - \left(1 - \frac{\alpha}{\rho} + \frac{\alpha^2}{2\rho^2}\right)b^2\right\}dx^2 - \left(1 + \frac{\alpha}{\rho}\right)\left(dy^2 + dz^2\right)$$

In the first and third term we replace $a^2$ by $1+b^2$ and delete in the first the quantity $\alpha^2 b^2/(2\rho^2)$, which is of third order; in the third we keep only the terms of the order 0 and 1 inside the curly brackets. In the second term we replace $a$ by 1 and in all terms $b$ by $v$. As a result, we obtain

$$ds^2 = \left(1 - \frac{\alpha}{\rho} + \frac{\alpha^2}{2\rho^2} - \frac{\alpha^2}{2\rho^2}v^2\right)dt^2 + \frac{4\alpha}{\rho}vdxdt - \left(1 + \frac{\alpha}{\rho}\right)(dx^2 + dy^2 + dz^2). \quad (68)$$

with $r = \sqrt{x^2 + y^2 + z^2}$ $\rho^2$ becomes $\rho^2 = r^2 + \tau^2 - t^2$ ...................(69)

We now observe everything at the moment $t = 0$. In the coordinate system $x,y,z,t$ we have at that moment in the origin $x = y = z = 0$ a field generating center, that moves along the positive $x$-axis at velocity $v$ and having the shape of an ellipsoid with the axis of symmetry being the shortest axis (the axes ratio is $1/a$). We demand the field in an arbitrary point $(x,y,z)$ at the moment $t = 0$. For that we shall take for $\rho$ in (68) the value, that follows from (69), namely

$$\rho^2 = r^2 + \tau^2 = r^2 + b^2 t^2,$$

because for $t = 0$, according to (67), $\tau = -bx$. Therefore,

$$\frac{1}{\rho} = \frac{1}{r}\left(1 - \frac{v^2 x^2}{2r^2}\right),$$

which is accurate up to the first order, and transforms (68) into

$$ds^2 = \left(1 - \frac{\alpha}{r} + \frac{\alpha^2}{2r^2} + \frac{\alpha^2 x^2}{2r^2}v^2 - \frac{2\alpha}{r}v^2\right)dt^2 + \frac{4\alpha}{\rho}vdxdt -$$

$$-\left(1 + \frac{\alpha}{\rho}\right)\left(dx^2 + dy^2 + dz^2\right) \qquad \text{..........(70)}$$

In this formula represents $\alpha/r$ in $g_{11}, g_{22}, g_{33}$ and $g_{44}$ the field in first order, the coefficient of $dxdt$ is exclusively due to motion and is of the order $1\tfrac{1}{2}$. The second order terms in $g_{11}, g_{22}$ and $g_{33}$ are ignored but are kept in $g_{44}$. The term $\alpha^2/(2r^3)$ is also present in the case of a center at rest and can conceivably originate from the interaction of the fields from the different parts of the center (in the case of two centers the product of the first order terms, caused by each center, would also appear). The term $-2\alpha v^2/r$ can be attributed partially to the changes in the values of the $T$'s as a result of the motion, and this term is also with the term $\alpha v^2 x^2/(2r^3)$ attributable to the finite propagation speed of gravity.

### § 2. Calculation of the field.

52. We have in mind a number of bodies moving in such a way, that the squares of the velocity components for each of the constituting points are of first order. Those components may be called $\dot{x}_1, \dot{x}_2, \dot{x}_3$; We take

$$\sum_{(i)} \dot{x}_i^2 = v^2.$$

We will substitute the following relations into the equations of gravity, (15),

$$g_{ij} = \alpha_{ij} + \alpha\beta_{ij} + \kappa^2\gamma_{ij}\,(i \neq 4, j \neq 4),\ g_{i4} = \alpha_{i4} + x^{3/2}\sigma_{i4}\,(i \neq 4),$$

$$g_{44} = \alpha_{44} + \kappa^{3/2}\beta_{44} + \kappa^2\gamma_{44},$$

$$g^{ij} = \alpha^{ij} + \kappa\beta^{ij} + \kappa^2\gamma^{ij},\ g^{i4} = \alpha^{i4} + \kappa^{3/2}\gamma^{i4},\ g^{44} = \alpha^{44} + \kappa\beta^{44} + \kappa^2\gamma^{44}.$$

The quantities $\beta$, $\sigma$ and $\gamma$ are of the order 0, just like the $\alpha$ quantities (see 48).

The term $\kappa^{3/2}\sigma_{i4}$ only contains the factor $\kappa$, as will become clear later, but, because it also contains a velocity component as factor, it is of the order $1\frac{1}{2}$. We write it as $\kappa^{3/2}\sigma_{i4}$, because this notation reminds us that it is actually of the order $1\frac{1}{2}$.

The expressions above are contained in the general expressions

$$g_{ij} = \alpha_{ij} + \kappa\beta_{ij} + \kappa^{3/2}\sigma_{ij} + \kappa^2\gamma_{ij},$$

$$g^{ij} = \alpha^{ij} + \kappa\beta^{ij} + \kappa^{3/2}\sigma^{ij} + \kappa^2\gamma^{ij},$$

which we can use instead of the formulas above, if we only assign $\sigma_{ij}$ and $\sigma^{ij}$ the value 0, when neither $i$ nor $j$ is equal to 4, and give $\beta_{ij}, \beta^{ij}, \gamma_{ij}, \gamma^{ij}$ the value 0 if only one of the indices is equal to 4.

When we now substitute the expressions above into the equations of gravity, it happens that some of the terms must be differentiated once or twice with respect to the time $t = x_4$.

53. We can form from the quantities $\beta_{ij}, \sigma_{ij}$ and $\gamma_{ij}$ the symbols

$$\begin{bmatrix} i\,j \\ l \end{bmatrix}_\beta,\ \begin{bmatrix} i\,j \\ l \end{bmatrix}_\sigma,\ \begin{bmatrix} i\,j \\ l \end{bmatrix}_\gamma$$

in the same way as the symbols of CHRISTOFFEL of the first kind (see 8) are built from the quantities $g_{ij}$, so that, for example, we have

$$\left[\begin{matrix}i\ j\\l\end{matrix}\right]_\beta = \tfrac{1}{2}\left(\frac{\partial^2 \beta_{il}}{\partial x_j} + \frac{\partial^2 \beta_{jl}}{\partial x_i} - \frac{\partial^2 \beta_{ij}}{\partial x_l}\right).$$

With this notation we have

$$\left\{\begin{matrix}i\ j\\l\end{matrix}\right\} = \sum_m g^{lm}\left[\begin{matrix}i\ j\\m\end{matrix}\right] =$$

$$\sum_m \left(\alpha^{lm} + \kappa\beta^{lm} + \kappa^{3/2}\sigma^{lm} + \kappa^2\gamma^{lm}\right)\left(\kappa\left[\begin{matrix}i\ j\\m\end{matrix}\right]_\beta + \kappa^{3/2}\left[\begin{matrix}i\ j\\m\end{matrix}\right]_\sigma + \kappa^2\left[\begin{matrix}i\ j\\m\end{matrix}\right]_\gamma\right)$$

and so, if we ignore terms of higher order than two,

$$\left\{\begin{matrix}i\ j\\l\end{matrix}\right\} = \kappa\alpha^{ll}\left[\begin{matrix}i\ j\\l\end{matrix}\right]_\beta + \kappa^{3/2}\alpha^{ll}\left[\begin{matrix}i\ j\\l\end{matrix}\right]_\sigma + \kappa^2\sum_m \beta^{lm}\left[\begin{matrix}i\ j\\m\end{matrix}\right]_\beta + \kappa^2\alpha^{ll}\left[\begin{matrix}i\ j\\l\end{matrix}\right]_\gamma \quad (71)$$

This expansion of the symbol of CHRISTOFFEL of the second kind must now be substituted in (15).

    54. We start with the calculation of the first order terms in (15), which equation we multiply by 2, so that it reads

$$2G_{ij} = -\kappa\left(2T_{ij} - g_{ij}T\right) \quad \text{...............................................................(72)}$$

We remember that, according to 9, we have

$$2G_{ij} = 2\sum_l \left( \frac{\partial \begin{Bmatrix} i\,l \\ l \end{Bmatrix}}{\partial x_j} - \frac{\partial \begin{Bmatrix} i\,j \\ l \end{Bmatrix}}{\partial x_l} \right) + 2\sum_{lm} \left( \begin{Bmatrix} i\,l \\ m \end{Bmatrix}\begin{Bmatrix} m\,j \\ l \end{Bmatrix} - \begin{Bmatrix} i\,l \\ m \end{Bmatrix}\begin{Bmatrix} m\,l \\ l \end{Bmatrix} \right)$$

This last expression exists, as one can see, of two parts, of which the first contains two derivatives of the $g$'s and the second products of derivates. For the time being we restrict ourselves to terms of the first order and can, therefore, ignore the second part of $2G_{ij}$; in the first part we only need to substitute the first term of (71), as a result we find

$$2\kappa \sum_l \alpha^{ll} \left( \frac{\partial}{\partial x_j}\left[ \begin{matrix} i\,l \\ l \end{matrix} \right]_\beta - \frac{\partial}{\partial x_l}\left[ \begin{matrix} i\,j \\ l \end{matrix} \right]_\beta \right) = \kappa \sum_l \left( \frac{\partial^2 \beta_{ll}}{\partial x_j \partial x_i} + \frac{\partial^2 \beta_{ij}}{\partial x_l^2} - \frac{\partial^2 \beta_{il}}{\partial x_l \partial x_j} - \frac{\partial^2 \beta_{jl}}{\partial x_l \partial x_i} \right)$$

or, if we keep the indices at 4 separately,

$$-\kappa \sum_{(l)} \left( \frac{\partial^2 \beta_{ij}}{\partial x_l^2} - \frac{\partial^2 \beta_{il}}{\partial x_l \partial x_j} - \frac{\partial^2 \beta_{jl}}{\partial x_l \partial x_i} \right) + \kappa \sum_l \alpha^{ll} \frac{\partial^2 \beta_{il}}{\partial x_l \partial x_j} + \kappa \left( \frac{\partial^2 \beta_{ij}}{\partial x_4^2} - \frac{\partial^2 \beta_{i4}}{\partial x_4 \partial x_j} - \frac{\partial^2 \beta_{j4}}{\partial x_4 \partial x_i} \right) \quad ....(73)$$

in which the last term is at least of the order two.

The terms of the first order become

$$\text{for } i \neq 4, j \neq 4 : -\kappa \sum_{(l)} \left( \frac{\partial^2 \beta_{ij}}{\partial x_l^2} - \frac{\partial^2 \beta_{il}}{\partial x_l \partial x_j} - \frac{\partial^2 \beta_{jl}}{\partial x_l \partial x_i} \right) + \kappa \sum_l \alpha^{ll} \frac{\partial^2 \beta_{ll}}{\partial x_j \partial x_i},$$

for $i \neq 4, j = 4$ : zero ,

$$\text{for } i = 4, j = 4 : -\kappa \sum_{(l)} \frac{\partial^2 \beta_{ij}}{\partial x_l^2} = -\kappa \Delta \beta_{44}.$$

The quantities $T_{ij}$, which appear in the second term of (72), are related to the tension-energy components $\sqrt{-g}T_i^j$ of the material, which were already discussed in 13 and which we, like EINSTEIN, represent as $\mathfrak{I}_i^j$. Namely

$$T_{ij} = \frac{1}{\sqrt{-g}}\sum_i g_{ij}\mathfrak{I}_i^j,$$

$$\mathfrak{I}_i^j = \sqrt{-g}\sum_l g^{ij}T_{il}.$$

The quantities $\mathfrak{I}_i^j (i \neq 4, j \neq 4)$ are the stress components, $-\mathfrak{I}_1^4, -\mathfrak{I}_2^4, -\mathfrak{I}_3^4$ the components of the amount of motion per unit of volume, $\mathfrak{I}_4^1, \mathfrak{I}_4^2, \mathfrak{I}_4^3$ those of the energy current, and $\mathfrak{I}_4^4$ is the energy per unit of volume. Between these quantities exist the relations expressed by

$$T_{ij} = T_{ji}.$$

To evaluate the order of magnitude we keep in mind what is said about the $g$'s in this regard. From there it follows that $-g$ deviates little from -1. The energy density $\mathfrak{I}_4^4$ will now be of the order 0. From the formulas above, the same is true about $T_{44}$ and also for the scalar quantity $T$ as follows from

$$T = \sum_{ij} g^{ij}T_{ij} = \frac{1}{\sqrt{-g}}\sum_i \mathfrak{I}_i^i.$$

If we now assume, that there are no other relevant amounts of motion than those, which are due to the velocity of material, then $\mathfrak{I}_4^1, \mathfrak{I}_4^2, \mathfrak{I}_4^3$ contain each a component of the velocity as a factor and are, therefore, of the order $\frac{1}{2}$. We will also assume that the tension components are due to gravity itself, or, at any rate, that no tension components are present, which are significantly larger than those caused by gravity. In that case the quantities $\mathfrak{I}_i^j$ for $i \neq 4$, $j \neq 4$ are of the first order.

From the relations between $T_{ij}$ and $\mathfrak{I}_i^j$ mentioned above, it follows now, that also $T_{ij}$ for $i \neq 4$, $j \neq 4$ is of first order, that $T_{i4} = T_{4i}$ is of the order $\frac{1}{2}$ and that $\mathfrak{I}_4^i$ (the components of the energy current) is also of the order $\frac{1}{2}$. If one only considers terms of the order 0, then we have

$$T = T_{44} = \mathfrak{I}_4^4,$$

because the differences between these three quantities with respect to each other, are of the order 1.

We consider $\rho$ to be of order 0, which differs from $T$, $T_{44}$ and $\mathfrak{I}_4^4$ only with an amount of the order 1. We note, hereby, that this definition does not determine the value of $\rho$ *exactly*. This indefiniteness is not relevant in the first part of the following calculations, but must be taken into account later.

For the first order terms in (72) we now find

$$\left. \begin{array}{c} \sum_{(l)} \left( \dfrac{\partial^2 \beta_{ij}}{\partial x_l^2} - \dfrac{\partial^2 \beta_{il}}{\partial x_l \partial x_j} - \dfrac{\partial^2 \beta_{jl}}{\partial x_l \partial x_i} - \dfrac{\partial^2 \beta_{jl}}{\partial x_l \partial x_i} \right) - \sum_l \alpha^{ll} \dfrac{\partial^2 \beta_{ll}}{\partial x_j \partial x_i} = \delta_{ij}\rho, (i \neq 4, j \neq 4) \\ \Delta \beta_{44} = \rho \end{array} \right\} \quad (74)$$

Here in is $\delta_{ij} = 1$ for $i = j$ and $\delta_{ij} = 0$ for $i \neq j$.

55. These equations can be satisfied in an infinite number of ways. The following solution is very simple

$$\beta_{ij} = 0 \, (i \neq j), \beta_{11} = \beta_{22} = \beta_{33} = \beta_{44} = \beta, \quad \text{...............................(75)}$$

in which the function $\beta$ satisfies

$$\Delta \beta = \rho \quad \text{...................................................(76)}$$

However, this solution is certainly not the only one. Because, if $\varphi$ is an arbitrary function of $x_1, x_2$ and $x_3$, then always

$$\beta_{ij} = \frac{\partial^2 \varphi}{\partial x_i \, \partial x_j} \quad \text{...................................................(77)}$$

is a solution of the system, that one obtains from (74), by taking $\rho = 0$; one can, therefore, add the solution (77) to (75) and get another valid solution. One obtains, this way, an unlimited number or new solutions.

The solution (75) is not only simple, but also has the additional advantage that $\kappa \beta_{44}$ for a single spherical center reduces to $-\alpha/\rho$ (a constant), which is of the same from as the gravitational potential in NEWTON's theory. However, it does not arise by ignoring terms of higher order than two in (28), which represents a spherical center at rest. Because, if one writes for (28)

$$ds^2 = \left(1 - \frac{\alpha}{r}\right) dt^2 - \left(\frac{1}{1 - \frac{\alpha}{r}} - 1\right) dr^2 - \left(dr^2 + r^2 d\vartheta^2 + r^2 \sin^2 \vartheta d\varphi^2\right),$$

then one can easily introduce Cartesian coordinates and finds, because of the identities

$$dr^2 + r^2 d\vartheta^2 + r^2 \sin^2 \vartheta d\varphi^2 = dx_1^2 + dx_2^2 + dx_3^2, \quad dr = \frac{x_1}{r} dx_1 + \frac{x_2}{r} dx_2 + \frac{x_3}{r} dx_3$$

that

$$ds^2 = \left(1 - \frac{\alpha}{r}\right)dt^2 - \frac{\alpha}{r-\alpha}\left(\frac{x_1}{r}dx_1 + \frac{x_2}{r}dx_2 + \frac{x_3}{r}dx_3\right)^2 - \left(dx_1^2 + dx_2^2 + dx_3^2\right).$$

After expanding this in a series of powers of $\alpha$, one finds

$$\beta_{44} = \frac{\alpha}{\kappa r}, \quad \beta_{ij} = \frac{\alpha}{\kappa r}\frac{x_i x_j}{r^2}(i \neq 4, j \neq 4).$$

These functions of $x_1, x_2, x_3$ form a solution of (74), but are different from (75); They originate from it by adding $\partial^2 \varphi / \partial x_i \partial x_j$, if $\kappa\varphi$ is equal to $\alpha r$.

However, we do obtain the solution (75) without adding (77), if we start from (31). In that case we have (equation 51)

$$ds^2 = \frac{\left(1 - \dfrac{\alpha}{4\rho}\right)^2}{\left(1 + \dfrac{\alpha}{4\rho}\right)^2} dt^2 - \left(1 + \frac{\alpha}{4\rho}\right)^4 \left(dx_1^2 + dx_2^2 + dx_3^2\right)$$

and one finds

$$\kappa\beta_{11} = \kappa\beta_{22} = \kappa\beta_{33} = \kappa\beta_{44} = -\frac{\alpha}{\rho}, \beta_{ij} = 0 \text{ if } i \neq j,$$

which is the solution (75) for a spherical center.

56. We will now set up the differential equations for the quantities $\sigma$ and $\gamma$. Substituting the solution (75) into (73) and deleting the terms of the first order, yields

$$\text{for } i \neq 4, j \neq 4 : \delta_{ij}\kappa\frac{\partial^2 \beta}{\partial x_4^2},$$

$$\text{for } i \neq 4, j = 4 : -2\kappa\frac{\partial^2 \beta}{\partial x_4 \partial x_i},$$

$$\text{for } i = j = 4 : -3\kappa\frac{\partial^2 \beta}{\partial x_4^2}.$$

So this is, after deleting the terms of the first order, the result of substituting the first term of (71) in the first part of $2G_{ij}$. We must also substitute the other terms of (71) in the first part of $2G_{ij}$ and finally also the first term of (71) into the second part of $2G_{ij}$.

If we now substitute the second term of (71) into the first part of $2G_{ij}$, then we find an expression, which differs from (73) only in so far, that everywhere one finds $\sigma$ in place of $\gamma$ and that $\kappa$ has there the power $\frac{3}{2}$. Because $\sigma_{ij}$ differs from 0 only when one of the two indices is equal to 4 and the other is equal to 1, 2 or 3, many terms drop out and what remains is

$$\text{for } i \neq 4, j \neq 4 : -\kappa^{\frac{3}{2}}\left(\frac{\partial^2 \sigma_{i4}}{\partial x_4 \partial x_j} + \frac{\partial^2 \sigma_{j4}}{\partial x_4 \partial x_i}\right), \text{ (of the order 2)}$$

$$\text{for } i \neq 4, j = 4 : \kappa^{\frac{3}{2}}\sum_{(l)}\left(\frac{\partial^2 \sigma_{4l}}{\partial x_l \partial x_i} - \frac{\partial^2 \sigma_{i4}}{\partial x_l^2}\right), \text{ (of the order } 1\tfrac{1}{2}\text{)}$$

$$\text{for } i = j == 4 : 2\kappa^{\frac{3}{2}} \frac{\partial^2 \sigma_{4l}}{\partial x_4 \partial x_l}. \text{ (of the order 2)}$$

It is necessary to substitute now also the third and the fourth term in the expansion of (71) into the first part of $2G_{ij}$. We start with the fourth term, because we can thereby also utilize the expression (73), replacing $\beta$ by $\gamma$ and $\kappa$ by $\kappa^2$. We find

$$\text{for } i \neq 4, j \neq 4 : -\kappa^2 \sum_{(l)} \left( \frac{\partial^2 \gamma_{ij}}{\partial x_l^2} - \frac{\partial^2 \gamma_{il}}{\partial x_l \partial x_j} - \frac{\partial^2 \gamma_{jl}}{\partial x_l \partial x_i} \right) + \kappa^2 \sum_l \alpha^{ll} \frac{\partial^2 \gamma_{ll}}{\partial x_i \partial x_j},$$

$$\text{for } i \neq 4, j = 4 : \text{ zero,}$$

$$\text{for } i = j = 4 : -\kappa^2 \sum_{(l)} \frac{\partial^2 \gamma_{44}}{\partial x_l^2} = -\kappa^2 \Delta \gamma_{44},$$

whereby terms of higher order than the second are ignored.

Before we now substitute the third term of (71) into the first part of $2G_{ij}$, we realize that because of $\beta^{ij} = 0 \ (i \neq j)$ and $\beta^{11} = \beta^{22} = \beta^{33} = \beta^{44} = -\beta$, that third term transforms into

$$-\kappa^2 \beta \begin{bmatrix} i j \\ l \end{bmatrix}_\beta.$$

As a result we obtain

$$-2\kappa^2 \sum_l \left\{ \frac{\partial}{\partial \dot{x}_j} \beta \begin{bmatrix} i\ l \\ l \end{bmatrix}_\beta - \frac{\partial}{\partial x_l} \beta \begin{bmatrix} i\ j \\ l \end{bmatrix}_\beta \right\} =$$

$$-4\kappa^2 \frac{\partial}{\partial \dot{x}_j}\left(\beta \frac{\partial \beta}{\partial x_i}\right) + \kappa^2 \sum_{(l)} \frac{\partial}{\partial \dot{x}_l}\left(\beta\left[\frac{\partial \beta_{il}}{\partial x_i} + \frac{\partial \beta_{jl}}{\partial x_i} - \frac{\partial \beta_{ij}}{\partial x_l}\right]\right) =$$

$$-2\kappa^2 \frac{\partial}{\partial x_j}\left(\beta \frac{\partial \beta}{\partial x_i}\right) - \delta_{ij}\kappa^2 \sum_{(l)} \frac{\partial}{\partial x_j}\left(\beta \frac{\partial \beta}{\partial x_i}\right),$$

in which terms of higher that the second order are ignored,

We obtain from here

$$\text{for } i \neq 4, j \neq 4 : -2\kappa^2 \frac{\partial}{\partial x_j}\left(\beta \frac{\partial \beta}{\partial x_i}\right) - \delta_{ij}\kappa^2 \sum_{(l)} \frac{\partial}{\partial x}\left(\beta \frac{\partial \beta}{\partial x_l}\right),$$

for $i \neq 4, j = 4$ : zero,

$$\text{for } i = j = 4 : -\kappa^2 \sum_{(l)} \frac{\partial}{\partial x_l}\left(\beta \frac{\partial \beta}{\partial x_l}\right),$$

in which we made use of the equality

$$\frac{\partial}{\partial x_i}\left(\beta \frac{\partial \beta}{\partial x_j}\right) = \frac{\partial}{\partial x_j}\left(\beta \frac{\partial \beta}{\partial x_i}\right)$$

and terms of higher order than the second are ignored.

Finally, we must still calculate the second part of $2G_{ij}$. Because we ignore terms of order higher than the second, we must only use the first term in the expansion in equation (71). As a result, the second part of $2G_{ij}$ becomes

$$2\kappa^2 \sum_{lm} \alpha^{ll} \alpha^{mm} \left( \begin{bmatrix} i\ l \\ m \end{bmatrix} \begin{bmatrix} j\ m \\ l \end{bmatrix} - \begin{bmatrix} i\ j \\ m \end{bmatrix} \begin{bmatrix} m\ l \\ l \end{bmatrix} \right) =$$

$$\kappa^2 \sum_{lm} \alpha^{ll} \alpha^{mm} \left\{ \frac{\partial \beta_{im}}{\partial x_l} \frac{\partial \beta_{jl}}{\partial x_m} - \frac{\partial \beta_{im}}{\partial x_l} \frac{\partial \beta_{jm}}{\partial x_l} + \tfrac{1}{2} \frac{\partial \beta_{lm}}{\partial x_i} \frac{\partial \beta_{lm}}{\partial x_j} - \tfrac{1}{2} \left( \frac{\partial \beta_{im}}{\partial x_j} + \frac{\partial \beta_{jm}}{\partial x_i} - \frac{\partial \beta_{il}}{\partial x_m} \right) \frac{\partial \beta_{ll}}{\partial x_m} \right\},$$

whereby in some terms the indices $l$ and $m$ have been interchanged. Because $\beta_{ij} = 0$ for two equal indices and $\beta_{ij} = \beta$ for two identical indices, this becomes equal to:

$$2\kappa^2 \alpha^{ii} \alpha^{jj} \frac{\partial \beta}{\partial x_i} \frac{\partial \beta}{\partial x_j} - \kappa^2 \alpha^{ij} \sum_l \alpha^{ll} \left( \frac{\partial \beta}{\partial x_l} \right)^2 + 2\kappa^2 \frac{\partial \beta}{\partial x_i} \frac{\partial \beta}{\partial x_j} -$$

$$- \tfrac{1}{2} \kappa^2 \left( \sum_l \alpha^{ll} \right) \left\{ (\alpha^{ii} + \alpha^{jj}) \frac{\partial \beta}{\partial x_i} \frac{\partial \beta}{\partial x_j} - \delta_{ij} \sum_m \alpha^{mm} \left( \frac{\partial \beta}{\partial x_m} \right)^2 \right\}.$$

If we now neglect the terms of higher order than two (this are the terms in which differentiation to $x_4$ takes place), then, because of $\sum_l \alpha^{ll} = -2$, the expression above is transformed into:

$$\kappa^2 \frac{\partial \beta}{\partial x_i} \frac{\partial \beta}{\partial x_j} + \kappa^2 \left( \delta_{ij} + \alpha^{ij} \right) \sum_{(l)} \left( \frac{\partial \beta}{\partial x_l} \right)^2.$$

The result of the substitution of (71) into the second part of $2G_{ij}$ is, therefore,

$$\text{for } i \neq 4, j \neq 4: \quad \kappa^2 \frac{\partial \beta}{\partial x_j} \frac{\partial \beta}{\partial x_j},$$

$$\text{for } i \neq 4, j = 4: \quad \text{zero},$$

$$\text{for } i = j = 4: \quad 2\kappa^2 \sum_{(l)} \left(\frac{\partial \beta}{\partial x_l}\right)^2.$$

In this way we find for $2G_{ij}$

$$\text{for } i \neq 4, j \neq 4: -\kappa^2 \sum_{(l)} \left(\frac{\partial^2 \gamma_{lj}}{\partial x_l^2} - \frac{\partial^2 \gamma_{il}}{\partial x_l \partial x_j} - \frac{\partial^2 \gamma_{jl}}{\partial x_l \partial x_i}\right) + \kappa^2 \sum_l \alpha^{ll} \frac{\partial^2 \gamma_{ll}}{\partial x_i \partial x_j} + \delta_{ij} \kappa \frac{\partial^2 \beta}{\partial x_4^2}$$

$$-\kappa^{\frac{3}{2}} \left(\frac{\partial^2 \sigma_{i4}}{\partial x_4 \partial x_j} + \frac{\partial^2 \sigma_{j4}}{\partial x_4 \partial x_i}\right) - 2\kappa^2 \frac{\partial}{\partial x_j}\left(\beta \frac{\partial \beta}{\partial x_i}\right) - \delta_{ij} \kappa^2 \sum_l \frac{\partial}{\partial x_l}\left(\beta \frac{\partial \beta}{\partial x_l}\right) + \kappa^2 \frac{\partial \beta}{\partial x_i} \frac{\partial \beta}{\partial x_j},$$

$$\text{for } i \neq 4, j = 4: \kappa^{\frac{3}{2}} \sum_{(l)} \left(\frac{\partial^2 \sigma_{4l}}{\partial x_l \partial x_i} - \frac{\partial^2 \sigma_{l4}}{\partial x_l^2}\right) - 2\kappa \frac{\partial^2 \beta}{\partial x_4 \partial x_l},$$

$$\text{for } i = j = 4: -\kappa^2 \Delta \gamma_{44} - 3\kappa \frac{\partial^2 \beta}{\partial x_4^2} + 2\kappa^{\frac{3}{2}} \sum_{(l)} \frac{\partial^2 \sigma_{4l}}{\partial x_4 \partial x_l}, -2\kappa^2 \sum_{(l)} \frac{\partial}{\partial x_l}\left(\beta \frac{\partial \beta}{\partial x_l}\right) + 2\kappa^2 \sum_{(l)} \left(\frac{\partial \beta}{\partial x_l}\right)^2.$$

57. The expressions thus calculated form each time the first term of the ten equations of gravity. If the second term of each of these equations is known, than one can start calculating $\sigma_{41}, \sigma_{42}$ and $\sigma_{43}$ from the triad, which arises with $i = 4$ $j = 1,2,3$. Thereafter $\gamma_{44}$ can be calculated from the equation arising for $i = j = 4$ and, if one then, finally, substitutes the values obtained for $\gamma_{44}, \sigma_{14}, \sigma_{24}$ and $\sigma_{34}$ in the nine other equations (of which there are only six different ones), than one could calculate the other $\gamma$'s, namely, $\gamma_{11}, \gamma_{22}, \gamma_{33}, \gamma_{23}, \gamma_{31}$ and $\gamma_{12}$ ( $\gamma_{14}, \gamma_{24}, \gamma_{34}$ are equal to zero). This last step is not necessary, however, because those $\gamma$'s yield in $g_{ij}$ ($i \neq 4, j \neq 4$) terms of the second order, but in $L$ they form terms of the third order. The equations, which originate from (72) for $i \neq 4, j \neq 4$ drop out, therefore, and only the equations for $i \neq 4, j = 4$ ($i = 4, j \neq 4$) and $i = j = 4$ remain.

We will simplify the first term of the equation. We have

$$\sum_{(l)} \frac{\partial}{\partial x_l}\left(\beta \frac{\partial \beta}{\partial x_l}\right) = \Delta\left(\tfrac{1}{2}\beta^2\right)$$

and also, because of (76):

$$2\sum_{(l)} \left(\frac{\partial \beta}{\partial x_l}\right)^2 = \Delta(\beta^2) - 2\beta\Delta\beta = \Delta(\beta^2) - 2\rho\beta.$$

As a result, the first term becomes, with $t$ instead of $x_4$:

$$-\kappa^2 \Delta\left(\gamma_{44} - \tfrac{1}{2}\beta^2\right) - 2\kappa^2 \rho\beta - 3\kappa \frac{\partial^2 \beta}{\partial t^2} + 2\kappa^{\frac{3}{2}} \sum_{(l)} \frac{\partial^2 \sigma_{4l}}{\partial t \, \partial x_l}.$$

Let us now consider the second term of the equations of gravity (72). The quantities $T_{ij}$ only need to be calculated up to, and including, the first order. We then obtain (see 54),

$$T = \sum_{ij} g^{ij} T_{ij} = \sum_i g^{ii} T_{ii} = (1 - \kappa\beta)T_{44} - \sum_{(l)} T_{ll}.$$

For $i \neq 4, j = 4$ the second term of (72) becomes

$$-2\kappa T_{i4}$$

and for $i = j = 4$ it becomes

$$-2\kappa T_{44} + \kappa(1+\kappa\beta)\left\{(1-\kappa\beta)T_{44} - \sum_{(l)} T_{ll}\right\} = -\kappa T_{44} - \kappa\sum_{(l)} T_{ll} = -\kappa\rho - \kappa(T_{44} - \rho) - \kappa\sum_{(l)} T_{ll}.$$

We have in the first term ignored the expression, which were already taken into account in 54; we should also do this in the second term, so that this becomes

$$-\kappa(T_{44} - \rho) - \kappa\sum_{(l)} T_{ll}.$$

We have, therefore, the following four equations for the calculation of $\sigma_{14}$, $\sigma_{24}$, $\sigma_{34}$, and $\gamma_{44}$:

$$\kappa^{\frac{1}{2}}\sum_{(l)}\left(\frac{\partial^2 \sigma_{l4}}{\partial x_l \partial x_i} - \frac{\partial^2 \sigma_{i4}}{\partial x_l^2}\right) = 2\frac{\partial^2 \beta}{\partial t \partial x_i} - 2T_{i4}, (i=1,2,3) \quad \ldots\ldots\ldots(78)$$

$$\kappa^2\Delta\left(\gamma_{44} - \tfrac{1}{2}\beta^2\right) = -3\kappa\frac{\partial^2 \beta}{\partial t^2} + 2\kappa^{\frac{3}{2}}\sum_{(l)}\frac{\partial^2 \sigma_{l4}}{\partial t \partial x_l} - 2\kappa^2\rho\beta + \kappa(T_{44} - \rho) + \kappa\sum_{(l)} T_{ll} \quad ..(79)$$

Finally we want to come back to actual $g_{i4}$ and $g_{44}$ instead of $\sigma_{l4}$ and $\gamma_{44}$.: If we set in (78) and (79)

$$\kappa^2 \sigma_{i4} = g_{i4}, \quad \kappa^2 \gamma_{44} = g_{44} - 1 - \kappa\beta$$

and take (76) into account, we obtain

$$\sum_{(l)}\left(\frac{\partial^2 g_{l4}}{\partial x_l \partial x_i} - \frac{\partial^2 g_{i4}}{\partial x_l^2}\right) = 2\kappa\frac{\partial^2 \beta}{\partial t \partial x_i} - 2\kappa T_{i4}, \quad \ldots\ldots\ldots\ldots\ldots\ldots(80)$$

$$\Delta\left(g_{44} - \tfrac{1}{2}\kappa^2\beta^2\right) = -3\kappa\frac{\partial^2 \beta}{\partial t^2} + 2\sum_{(l)}\frac{\partial^2 g_{l4}}{\partial t \partial x_l} - 2\kappa^2\rho\beta + \kappa\sum_{(l)} T_{ll} \quad \ldots\ldots\ldots\ldots(81)$$

58. The equations (80) and (81) must now be solved. For that purpose it is first necessary to get to know the $T$ 's.

First of all, $\mathfrak{I}_j^4 (j=1,2,3)$ is a component of the amount of motion and is, therefore, accurate up to terms of the order $\frac{1}{2}$, equal to $\rho \dot{x}_j$. From

$$T_{i4} = \frac{1}{\sqrt{-g}} \sum_l g_{l4} \mathfrak{I}_i^l$$

follows then, if one takes in consideration what is said about $\mathfrak{I}_i^l$ for $i \neq 4, l \neq 4$ above, that one, if $i \neq 4$, can write

$$T_{i4} = -\rho \dot{x}_i, \quad\quad\quad\quad\quad\quad\quad\quad\quad\quad\quad\quad\quad\quad (82)$$

accurate up to terms of the order $\frac{1}{2}$, which is sufficient for (80).

In order to calculate $\sum_l T_{il}$ in (81), accurate up to terms of the first order, we will first perform a coordinate transformation, whereby $t$ is not involved and which in the point, where we will calculate $\sum_l T_{il}$, causes the direction of motion to coincide with the direction $dy_2 = dy_3 = 0$, while the speed becomes $v = \dot{y}_1$. Then we have

$$\sum_{(l)} T_{ll}' = \sum_{(lij)} \frac{\partial x_i}{\partial y_l} \frac{\partial x_j}{\partial y_l} T_{ij}, \quad T_{44}' = T_{44}$$

and because, if we assume in both systems the axes to be perpendicular to each other,

$$\sum_{(l)} \frac{\partial x_i}{\partial y_l} \frac{\partial x_j}{\partial y_l} = \delta_{ij},$$

we will have

$$\sum_{(l)} T_{ll}' = \sum_{(l)} T_{ll}$$

and, therefore, also

$$\sum_{l} T_{ll}' = \sum_{l} T_{ll}.$$

We are allowed, therefore, in (81) to regard $\sum_{l} T_{ll}$ as belonging to the $y_1, y_2, y_3, t$ coordinate system.

Now we perform a transformation, as is used in the special theory of relativity, by which we require the mass in the point under consideration to be at rest. Take

$$y_1' = ay_1 - bt, \; y_2' = y_2, \; y_3' = y_3, \; t' = at - by_1,$$

in which holds:

$$a^2 - b^2 = 1, \; a = \left(1 - v^2\right)^{-\frac{1}{2}}, \; b = v\left(1 - v^2\right)^{-\frac{1}{2}}$$

If $T_{ij}'$ refers to the system $y_1', y_2', y_3', t'$ and $T_{ij}$ to $y_1, y_2, y_3, t$, then from

$$\sum_{l} T_{ll} = \sum_{ijl} \frac{\partial y_i'}{\partial y_l} \frac{\partial y_j'}{\partial y_l} T_{ij}'$$

follows easily

$$\sum_l T_{ll} = \left(a^2 T_{11}' - 2ab T_{14}' + b^2 T_{44}'\right) + T_{22}' + T_{33}' +$$

$$+ \left(b^2 T_{11}' - 2ab T_{14}' + a^2 T_{44}'\right).$$

Now we have in the system $y_1', y_2', y_3', t'$

$$T_{14}' = \frac{1}{\sqrt{-g}} \sum_l g_{l4}' \Im_1^{l'}$$

and because $\Im_1^{4'}$ is equal to zero, one only needs to take $l = 1,2,3$; because $T_{14}'$ becomes of the order $\tfrac{5}{2}$ and $bT_{14}'$ of the order 3, one sees that this term can be dropped. Replacing $a^2$ by $1 + b^2$ yields

$$\sum_l T_{ll} = \sum_l T_{ll}' + 2b^2 T_{44}', \quad\text{...........................(83)}$$

in which we ignored $2b^2 T_{11}'$ as it is of second order.

We may now replace $2b^2 T_{44}'$ by $2v^2 \rho$. If we now substitute (82) in (80) and (83) in (81), we obtain

$$\sum_{(l)} \left(\frac{\partial^2 g_{l4}}{\partial x_l \partial x_i} - \frac{\partial^2 g_{i4}}{\partial x_l^2}\right) = 2\kappa \frac{\partial^2 \beta}{\partial t \partial x_i} + 2\kappa \rho \dot{x}_i, \quad\text{...........................(84)}$$

$$\Delta\left(g_{44} - \tfrac{1}{2}\kappa^2 \beta^2\right) - 3\kappa \frac{\partial^2 \beta}{\partial t^2} + 2\sum_{(l)} \frac{\partial^2 g_{l4}}{\partial t \partial x_l} - 2\kappa^2 \rho \beta + 2\kappa \rho v^2 + \kappa \sum_l T_{ll}' \quad\text{....(85)}$$

59. We will now solve (84) and (85). The following expression satisfies (84):

$$g_{i4} = 2\kappa \int \frac{\rho \dot{x}_i dS}{4\pi r} \quad \text{...(86)}$$

Herein is $r$ the distance (square-root of the sum of the squares of the differences between coordinates) of the point, where $g_{i4}$ must be calculated, to the point that relates to $\rho$ and $\dot{x}_i$ and that is situated in the element $dS$. One sees easily, that (86) satisfies (84). First, from (86) follows:

$$-\sum_{(l)} \frac{\partial^2 g_{i4}}{\partial x_l^2} = 2\kappa \rho \dot{x}_i,$$

so that it remains to prove, that the following relation holds

$$\sum_{(l)} \frac{\partial g_{l4}}{\partial x_l} = 2\kappa \frac{\partial \beta}{\partial t} \quad \text{...(87)}$$

We find first the following:

$$\sum_{(l)} \frac{\partial g_{l4}}{\partial x_l} = -2\kappa \int \frac{\rho dS}{4\pi} \sum_{(l)} \dot{x}_l \frac{\partial}{\partial x_l}\left(\frac{1}{r}\right), \quad \text{...(88)}$$

in which the differentiation $\frac{\partial}{\partial x_l}\left(\frac{1}{r}\right)$ must be done with respect to the coordinate $x_l$ of the element $dS$.

We imagine now, that the points of the element $dS$ participate in the motion of the material, so that, when $\rho$ were the density of a material with *constant* mass, $\rho dS$ would be independent of the time. This is not necessarily the case, but we will assume, that the change in $\rho dS$ will be of higher order than $\frac{1}{2}$ (see 62). Then we can write

$$2\kappa\frac{\partial \beta}{\partial t} = -2\kappa\frac{\partial}{\partial t}\int\frac{\rho dS}{4\pi} = -2\kappa\int\frac{\rho dS}{4\pi}\sum_{(l)}\dot{x}_l\frac{\partial}{\partial x_l}\left(\frac{1}{r}\right),$$

because terms of order higher than $1\tfrac{1}{2}$ can be left out. Therefore, is (87) satisfied as a result of (88), and, as a result, (86) solves (84).

From (87) follows

$$\sum_{(l)}\frac{\partial^2 g_{l4}}{\partial t\,\partial x} = 2\kappa\frac{\partial^2 \beta}{\partial t^2}$$

and with this (85) becomes

$$\Delta\left(g_{44} - \tfrac{1}{2}\kappa^2\beta^2\right) = \kappa\frac{\partial^2 \beta}{\partial t^2} - 2\kappa^2\rho\beta + 2\kappa\rho v^2 + \kappa\sum_{l}T_{ll}{'} \quad\text{..........(89)}$$

60. Before we solve (89), we first note, that we have

$$\Delta B = -2\kappa\beta,$$

if the following is true

$$B = \kappa\int\frac{\rho r dS}{4\pi} \quad\text{..........(90)}$$

In a point, around which one can draw a sphere, within which $\rho$ is equal to 0 everywhere, this is obvious, because one finds after executing the $\Delta$ operation under the integral sign:

$$\Delta B = 2\kappa\int\frac{\rho dS}{4\pi r},$$

which is according to (76) equal to $-2\kappa\beta$. In a point, that is situated inside the material, one first chooses a small sphere and proves, that

$$\frac{\partial B}{\partial x_i} = \kappa \int \frac{\rho dS}{4\pi} \frac{\partial r}{\partial x_i}$$

is correct, in the same way as one can prove in the theory of potentials, that

$$\frac{\partial \beta}{\partial x_i} = -\int \frac{\rho dS}{4\pi} \frac{\partial \left(\frac{1}{r}\right)}{\partial x_i}$$

holds.  The same analysis can be applied to the result again and then the proof is completed.  We find in this way

$$\kappa \frac{\partial^2 \beta}{\partial x^2} = \frac{\partial^2}{\partial t^2}(-\tfrac{1}{2}\Delta B) = \Delta\left(-\tfrac{1}{2}\frac{\partial^2 \beta}{\partial t^2}\right)$$

and, as a result, (89) becomes

$$\Delta\left(g_{44} - \tfrac{1}{2}\kappa^2 \beta^2 + \tfrac{1}{2}\kappa \frac{\partial^2 \beta}{\partial t^2}\right) = \kappa \sum_l T_{ll}' + 2\kappa\rho v^2 - 2\kappa^2 \rho\beta \quad \text{....(91)}$$

The solution of this equation is

$$g_{44} = 1 + \tfrac{1}{2}\kappa^2 \beta^2 - \tfrac{1}{2}\frac{\partial^2 B}{\partial t^2} - \kappa \int \frac{\sum_l T_{ll}'}{4\pi r} dS - 2\kappa \int \frac{\rho v^2}{4\pi r} dS + 2\kappa^2 \int \frac{\rho \beta dS}{4\pi r} \quad \text{...(92)}$$

The calculation of the field is herewith completed, in so far this is possible without knowing the nature and the distribution of the material.  In the next section we will apply this result to the calculation of the field from a number of spheres, a problem that is relevant for Astronomy.

## § 3. **Application to spherical bodies.**

61. We have in mind a number of bodies, of which the sizes are small with respect to the distances between them, and which are spherical, when they are at rest. In the mean time we assume that they are in motion in such a way, that one can assign the same velocity to each of the points of each body; which is for the first body $v_1$ with the components $\dot{x}_1, \dot{y}_1, \dot{z}_1$, and in general for the *i*-th body $v_i$ with components $\dot{x}_i, \dot{y}_i, \dot{z}_i$. We assume, in agreement with 52, that the quantities $\dot{x}_i, \dot{y}_i, \dot{z}_i$ are of the order $\tfrac{1}{2}$. We set

$$\int_{(i)} \frac{\rho dS}{4\pi r} = -\beta_i, \quad\quad\quad\quad (93)$$

in which the $(i)$ below the integral sign indicates, that one must integrate over the *i*-th body.

Then we have

$$\beta = \sum_i \beta_i.$$

Because $\dot{x}_i, \dot{y}_i, \dot{z}_i$ for all points of the *i*-th body are equally large, it follows from (86) that we have

$$g_{14} = 2\kappa \sum_i \int_{(i)} \frac{\rho \dot{x}_i dS}{4\pi r} = 2\kappa \sum_i \dot{x}_i \int_{(i)} \frac{\rho dS}{4\pi r} = -2\kappa \sum_i \beta_i \dot{x}_i.$$

Similar equations are true for $g_{24}$ and $g_{34}$. So, we have

$$g_{14} = -2\kappa \sum_i \beta_i \dot{x}_i, \quad g_{24} = -2\kappa \sum_i \beta_i \dot{x}_i, \quad g_{14} = -2\kappa \sum_i \beta_i \dot{z}_i \quad ......(94)$$

62. Now we come to the calculation of $g_{44}$. At the calculation of the terms of the second order in (92), we do not need to take into account with a deviation of the spherical shape, nor with the mutual interactions between the centers, not with the motion. Because the sizes of the bodies are small compared to the distances between them, we can consider $r$ in (90) to be a constant and, if we set

$$\kappa \int_{(i)} \rho \, dS = 8\pi k_i, \qquad (95)$$

we find

$$B = 2\sum_i k_i r_i,$$

if $r_i$ represents the distance from the center of $i$-th body to the point, where $B$ must be calculated.

From (93) follows

$$\kappa \beta_i = -\frac{2k_i}{r_i} \qquad (96)$$

and furthermore we have

$$2\kappa \int \frac{\rho v^2 dS}{4\pi r} = \sum_i \frac{4k_i}{r_i} v_i^2. \qquad (97)$$

In the last term of (92) we split $\beta$ inside the $i$-th body in two parts, so that it becomes

$$\beta = \beta_i + (\beta),$$

in which $(\beta)$ refers to all bodies, except the $i$-th body. We can now consider $(\beta)$ to be constant inside the $i$-th body and thus obtain

$$2\kappa^2 \int_{(i)} \frac{\rho(\beta) dS}{4\pi r} = 2\kappa^2 (\beta) \int_{(i)} \frac{\rho dS}{4\pi r} = 4\kappa (\beta) \frac{k_i}{r} = -8 \sum_{j \neq i} \frac{k_j}{r_{ij}},$$

in which $r_{ij}$ represents the distance from the center of the *i*-th body to that of the *j*-th body and *j* does not assume the value *i*.

From this equation and from (92), (6) and (97) follows

$$g_{44} = 1 - \kappa \sum_i \int_{(j)} \frac{\left(\sum_l T_{ll}' - 2\kappa\rho\beta_i\right)}{4\pi r} dS +$$

$$+2\left(\sum_i \frac{k_i}{r_i}\right)^2 - \sum_i k_i \frac{\partial^2 r_i}{\partial t^2} - 4\sum_i \frac{k_i}{r_i} v_i^2 - 8\sum_i \frac{k_i}{r_i} \sum_{j\neq i} \frac{k_j}{r_{ij}}. \quad \ldots(98)$$

Hereby we note, that, if, as we assumed above, $d(\rho dS)/dt$ is of higher order than $\frac{1}{2}$, according to (95) $dk_i/dt$ is of higher order than $1\frac{1}{2}$ and so $d^2 k_i/dt^2$ is of higher order than 2. In

$$\frac{d^2(k_i r_i)}{dt^2} = k_i \frac{\partial^2 r_i}{\partial t^2} + 2\frac{dk_i}{dt}\frac{\partial r_i}{\partial t} + \frac{d^2 k_i}{dt^2} r_i$$

we can, therefore, ignore the last two terms, which are of higher than the second order.

That $d(\rho dS)/dt$ is indeed of higher order than $\frac{1}{2}$ can be seen as follows. First of all, $d(dS)/dt$ is of order $1\frac{1}{2}$, and, even though $d(dS)/dt$ is not exactly equal to 0 as a result of the contraction (which depends on $v$), it is sufficient to show that $d\rho/dt$ is of higher order than $\frac{1}{2}$. Because $\mathfrak{I}_4^4$ differs from $\rho$ with a quantity of the first order, we only need to show, that $d\mathfrak{I}_4^4/dt$ is of higher order than $\frac{1}{2}$. We now apply on the system $y_1', y_2', y_3', t$ the transformation which we also used between equations (82) and (83) ; whereby we choose $a$ and $b$ in such a way, that at one moment the point under consideration is at rest, but we also apply the transformation to nearby moments. One has then

$$\mathfrak{I}_4^4 = a^2 \mathfrak{I}_4^4{}' + ab\left(\mathfrak{I}_4^1{}' - \mathfrak{I}_1^4{}'\right) - b^2 \mathfrak{I}_1^1{}'$$

and from here we see, that we only need to prove that $d\mathfrak{I}_4^4{}'/dt$ is of higher order than $\frac{1}{2}$. Instead, we may also prove this for $d\mathfrak{I}_4^4{}'/dt'$ and, finally, we can also consider

$$\frac{d}{dt'}\int \mathfrak{I}_4^4{}' dS',$$

in which the integral extends over the whole body at rest, at the moment under consideration. The quantity $d\left(\int \mathfrak{I}_4^4{}' dS'\right)/dt$ must be equal to the total energy current through the area minus the energy change per unit of time in the field of gravity present inside the area, because the integral represents the existing energy, and the change in position of the material during the time $dt'$ can be ignored. This last energy contains the factor $\kappa$ and also the energy current contains the factor $\kappa$, if we ignore energy currents other than that of gravity. With this, it is shown that $d(\rho dS)/dt$ is of higher order than $\frac{1}{2}$.

We still must calculate the second term In (98). It is clear, that the influence of the field of the other bodies on the values $T_{11}'$, $T_{22}'$, $T_{33}'$ and $T_{44}'$ in the *i*-th body is of the first order; but it is, due to the large distance compared to the size of the *i*-th body, very small. If we ignore this influence (which is for astronomical applications certainly allowed), and if we ignore also the change in shape of the *i*-th body due to the other bodies, then the integrand and the area of integration of the *i*-th term of the first sum in (98) are only determined by the *i*-th body.

First we will take the integral, which, because of the flattening extends necessarily over an ellipsoid, over a sphere, which is allowed, if we add a factor $1 - \frac{1}{2} v_i^2$. In (95) we can integrate over a sphere without further justification. For the second order term arising in the second term of (98) we can take $\sum_i v_i^2 k_i / r_i$ after multiplication with $-v_i^2/2$. We find then

$$g_{44} = 1 - \kappa \sum_i \int_{(j)} \frac{\left( \sum_l T_{ll}' - 2\kappa \rho \beta_i \right)}{4 \pi r} dS +$$

$$+ 2 \left( \sum_i \frac{k_i}{r_i} \right)^2 - \sum_i k_i \frac{\partial^2 r_i}{\partial t^2} - 3 \sum_i \frac{k_i}{r_i} v_i^2 - 8 \sum_i \frac{k_i}{r_i} \sum_{j \neq i} \frac{k_j}{r_{ij}} \quad \ldots\ldots(99)$$

In view of the large distance between the spheres compared to their radii, we can now write for the second term in (99)

$$- \sum_i \frac{2 k_i}{r_i},$$

if we take

$$8 \pi k_i = \int_{(i)} \left( \sum_{ll} T_{ll}' - 2 \kappa \rho \beta_i \right) dS,$$

and we are then also allowed to use this $k_i$ in the other terms of (99) instead of the quantity $k_i$, which is determined by (95). We do not know, however, whether this $k_i$ will remain constant for longer time. If we could specify quantities, which do have this property, then it would be preferable to use these in (99) instead of $k_i$. We must leave this question undecided and keep the form, that (99) gives for $g_{44}$.

From (94), (96), and from

$$g_{11} = g_{22} = g_{33} = -1 + \kappa\beta$$

follows finally

$$ds^2 = g_{44}dt^2 + 8\sum_i \frac{k_i}{r_i}(\dot{x}_i dx + \dot{y}_i dy + \dot{z}_i dz)dt -$$

$$-\left(1 + 2\sum_i \frac{k_i}{r_i}\right)(dx^2 + dy^2 + dz^2), \quad \ldots\ldots\ldots(100)$$

in which $g_{44}$ is given by (99).

63. We just want to apply this formula to the case of one single center moving uniformly along the $x$-axis. First we have

$$\frac{\partial^2 r}{\partial t^2} = \frac{v^2}{r} - \frac{x^2}{r^3}v^2.$$

If one writes further in (99) in the second term $\sum_l T_{ll}'$ in the place of $\rho$ and also in (93), whereby one can integrate over a sphere, then in the second term of (99) the quantity

$$\left\{1+2\kappa\int\frac{\left(\sum_l T_{ll}'\right)dS}{4\pi r}\right\}\sum_l T_{ll}'$$

arises under the integral sign. We can take this quantity for $\rho$ in (95) (whereby the integration is spherical); $k$ is then a constant, because the $\rho$ does not depend on the sate of motion now. The second term in (99) becomes

$$-\frac{2k}{r}$$

and, therefore, we find, because we can replace in the other terms the old $k$ by the new $k$,

$$ds^2 = \left(1 - \frac{2k}{r} + \frac{2k^2}{r^2} - 4v^2\frac{k}{r} + \frac{x^2}{r^2}v^2\frac{k}{r}\right)dt^2 + 8v\frac{k}{r}dxdt$$
$$- \left(1 + \frac{2k}{r}\right)(dx^2 + dy^2 + dz^2),$$

or, if we replace $k$ by $\tfrac{1}{2}\alpha$,

$$ds^2 = \left(1 - \frac{\alpha}{r} + \frac{\alpha^2}{2r^2} - 2v^2\frac{\alpha}{r} + \frac{x^2}{2r^2}v^2\frac{\alpha}{r}\right)dt^2 + 4v\frac{\alpha}{r}dxdt$$
$$- \left(1 + \frac{\alpha}{r}\right)(dx^2 + dy^2 + dz^2),$$

which is in perfect agreement with (70). Therewith is (100), which we set out to derive, verified for a special case.

---

## PROPOSITIONS.

------------

### I.
The so-called basic theorem of Algebra is not relevant for Algebra.

### II.
Analytical functions exist, for which none of the singularities are isolated, while they do not form a line or a plane; this is what, for example, A.R. FORSYTH has overlooked in his well-known text book on the theory of functions (chapter VII).

### III.
It is not only possible, as VOLTERRA has shown, to construct a function, of which the derivative in a certain interval cannot be integrated (a la RIEMANN), but there are also functions, of which the derivative cannot in any sub-interval be integrated.

### IV.
If in the complex plane a set of points is given in such a way, that each pair of its internal points can be connected by a chain of circles, for which each following pair intersects and for which the first one point of this pair and the last the other point incloses, then a unique analytical function exists, which is regular in each internal point of the set and cannot be continued outside this set.

## V.

There are no unique analytical functions $f(z)$, which satisfy the functional equation $f(z^n) = f(z)$, if $n$ is an integer >1 is.

## VI.

The axiom of M. PASCH (Vorlesungen über neuere Geometrie, page 21, Grundsatz IV) must be formulated in HILBERT's system of axiom's (Grundlagen der Geometrie, 3rd printing, axiom II 4) in such a way that in the plane of a triangle no line exists, which (without passing through one of the corners) intersects with only on of the sides.

## VII.

The question of the "curvature" of our space has lost its meaning.

## VIII.

The integral equation, with which HILBERT treats the theory of gasses for the case of completely elastic spherical molecules with the proposition of the usual formula for the number of collisions, has no other Eigenvalue other than 1.

## IX.

The theory of integral equations is only of little use for Physics.

## X.

When in the general theory of relativity it would become desirable to restrict one self to coordinate systems, which satisfy one single conditions (for example ($\sqrt{-g} = 1$), then this is not necessarily a reason to consider the attempt to construct the theory as h a failure.

### XI.

It cannot be justified, that at an isotherm the pressure, at which equilibrium between two phases can exist, is determined by the consideration, that $\int p\,dv$ along both isotherms must be equal.

### XII.

The secondary education in mathematics should be reformed.

———

(Translator's note: The appearance of propositions at Ph.D. theses is a Dutch tradition. The candidate is expected to be able to defend his thesis but also his propositions if challenged).

109